\documentclass{article}

\usepackage{arxiv}

\usepackage[utf8]{inputenc} 
\usepackage[T1]{fontenc}    
\usepackage{hyperref}       
\usepackage{url}            
\usepackage{booktabs}       
\usepackage{amsfonts}       
\usepackage{nicefrac}       
\usepackage{microtype}      
\usepackage{lipsum}
\usepackage{amsmath}
\usepackage{graphicx}

\usepackage{float}
\floatstyle{boxed} 
\restylefloat{figure}
\usepackage{caption}
\usepackage{subcaption}

\usepackage{amsmath}

\graphicspath{{Images/}}

\title{\textbf{An error reduced and uniform parameter approximation in fitting of B-spline curves to data points}}

\author{
   Debashis Mukherjee \\
   Graduate Student, \\
  Jadavpur University, Kolkata, India\\
  \textit{debashis\_mukherjee@yahoo.com}
}

\begin{document}
\maketitle
\begin{abstract} \normalsize

Approximating data points in three or higher dimension space based on cubic B-spline curve is presented.
Representations for planar curves, are merged and extended to the higher dimension.
The curve is fitted to the order of data points, or uniform parameter values are assumed for the points.   
Tangents are assumed at the data points, corresponding to the property used in cardinal splines, for shape preserving and visually pleasing fit.
Control points of piecewise continuous cubic bezier curves, meeting the boundary conditions of cardinal spline segments, are used for b-spline curve in corresponding coordinate planes.  
Approximation using error computed in the least square sense, based on a fraction of data points, is also presented.

\end{abstract}

\keywords{Data fitting \and B-splines \and Cardinal splines}

\vfill

\section{Introduction}

Parametric curve and surface approximation of B-splines, corresponding to parameter estimated of data points depending on chord length, centripetal, or base curves and surfaces parameterizations, based on numerical optimization using newton type methods, and minimization of error in the least square sense, is well studied ~\cite{ Borges,Ma,Sung,Park,Ravari}. Achieving of high accuracy in fitting to the constraints of the parameters, cause to increase the degree of the curve and surface. A high degree fit can lower the shape preserving and visually pleasing nature ~\cite{ Foley, Goodman, Fritsch, Martin, Hoscheck, Catmull}, and increases computational cost with the modelled curve and surface further in geometric design. In this context, we present algorithm for fitting a cubic b-spline curve to data points in three or higher dimensional space, with properties of continuity naturally upto $C^2$, and agrees with tangents at data points related to shape preserving and visually pleasing interpolation at projections in corresponding coordinate planes.

This paper is organized as follows. In Section 2 we discuss basic concepts and  terminologies used in this paper. Algorithms and rules for construction of curve fitting are discussed in Section 3. Related work is discussed in Section 4. Results of experimentation is presented in Section 5. Finally, this paper is concluded in Section 6.

\begin{figure*}[!b]
\centering{\includegraphics[width=10cm,height=10cm,keepaspectratio]{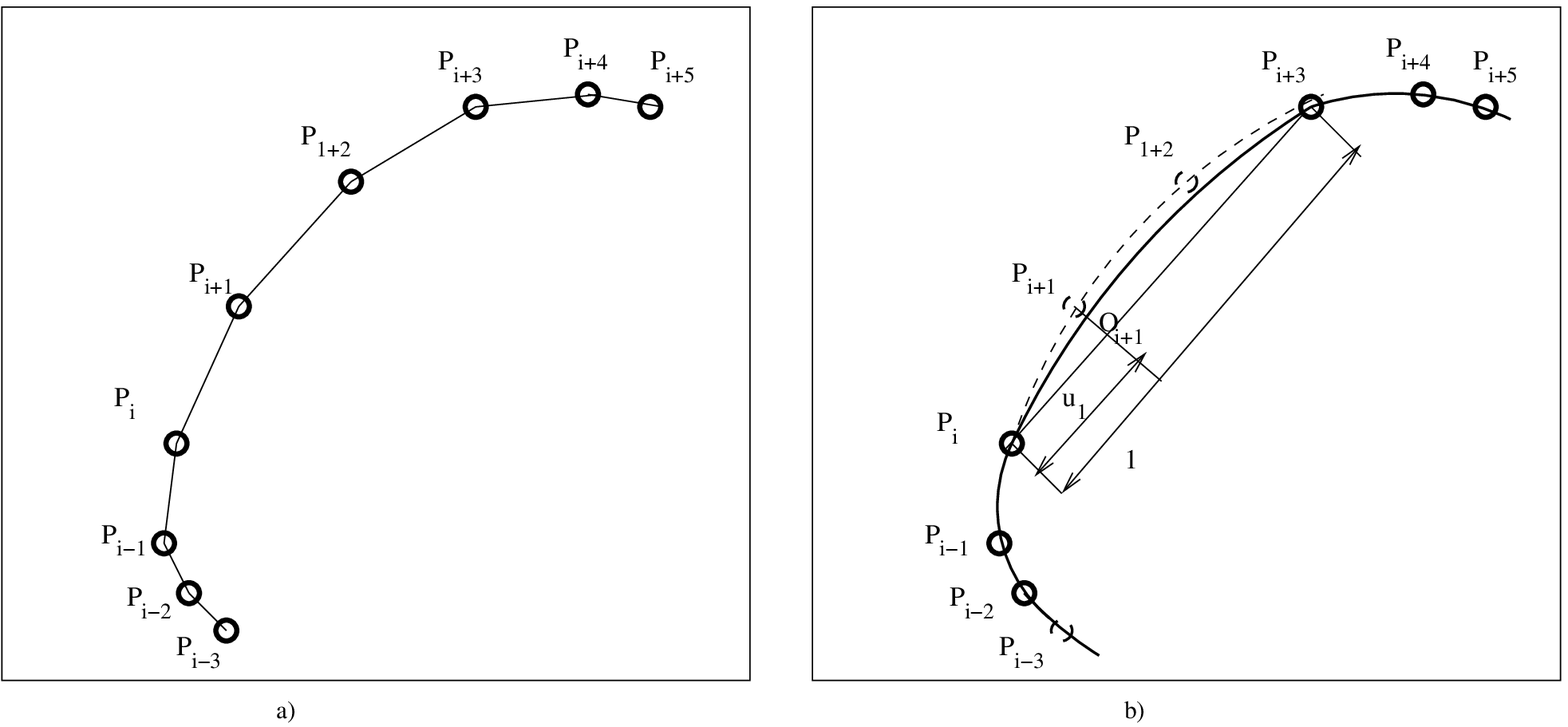}}
\caption{Dominant Points.}
\label{dpntfig}
\end{figure*}

\section{Basic concepts and terminologies}
\label{sec1_}

In this Section, we discuss basic concepts and terminologies, necessary in presenting our algorithm of B-spline curve fitting in Section 3.

\subsection{Cardinal Spline}
Cardinal spline is a cubic hermit class of spline. Hermit interpolation uses cubic polynomials to interpolate consecutive data points in pairs. An equality of the tangent values, of the individual segments ending and starting at every intermediate data point is ensured.   Thus, a piecewise cubic curve interpolating the data points, with $C^{1}$ continuity at the junctions is obtained. A cardinal spline, defines equality of tangent values at every intermediate data point, by specifying, the tangent at every point. The tangent, at every intermediate data point, is specified to be equal to a constant multiple of the vector, using the next point as the end point, and the previous point as the start point for the corresponding vectors. Catmull-Rom spline ~\cite{Catmull} is a specific cardinal spline that uses a fixed value of 0.5 for the constant.

For, a cardinal spline to pass through, a sequence ($p_0$, $p_1$, …, $p_n$), of ($n$+1) data  or control points, the tangent at $p_k$, $k\in$ $\mathbb{Z} \cap [0,n]$ is expressed as, $\tau$($p_{k+1}$ - $p_{k-1}$), where $\tau$ is the constant, and is considered to represent the tension parameter of the curve segment.  The parametric equation,  $\rho(u)$=$au^3$ + $bu^2$ + $cu$ + $d$, of a curve segment, for range  of parameter $0 \le u  \le 1 $, is defined, for the curve between control points $p_{k-1}$ to $p_k$. The geometry matrix, for the curve is derived using the control points,  $p_{k-2}$, $p_{k-1}$, $p_{k}$, and $p_{k+1}$. The boundary conditions, used are ,  $\rho$(0)=$p_{k-1}$, $\rho$(1)=$p_{k}$, $\rho^{'}$(0)=$\tau$($p_{k}$-$p_{k-2}$), and $\rho^{'}$(1)=$\tau$($p_{k+1}$-$p_{k-1}$), based on the coordinates and  tangent values, at control points, $p_{k-1}$, and $p_{k}$. The control points $p_0$, and $p_n$, are required to be repeated, in the sequence of control points, to generate points for the curve segments between control points $p_0$ to $p_1$, and $p_{n-1}$ to $p_n$, respectively. 
The parametric equation for the $k$th curve segment is expressed as,

\begin{equation}
\rho(u) = \begin{pmatrix}
u^3 & u^2 & u & 1 
\end{pmatrix}
\begin{pmatrix}
a \\
b \\
c  \\
d 
\end{pmatrix}
=
\begin{pmatrix}
u^3 & u^2 & u & 1 
\end{pmatrix}
\begin{pmatrix}
-\tau & (2 - \tau) & (\tau - 2) & \tau \\
2\tau & (\tau - 3) & (3 - 2\tau) & -\tau \\
-\tau  & 0  & \tau & 0  \\
0 & 1 & 0 & 0 
\end{pmatrix}
\begin{pmatrix}
p_{k-2} \\
p_{k-1} \\
p_k  \\
p_{k+1} 
\end{pmatrix}
\end{equation}

where, matrix M=$\begin{pmatrix}
-\tau & (2 - \tau) & (\tau - 2) & \tau \\
2\tau & (\tau - 3) & (3 - 2\tau) & -\tau \\
-\tau  & 0  & \tau & 0  \\
0 & 1 & 0 & 0 
\end{pmatrix}$, corresponds to the geometry matrix (GM) of the curve.



\subsection{Bezier Curve}
Bezier curve is an approximating spline, since the control points representing the geometric description of the curve are not required to be points on the curve.
Bezier curves are constructed using any number of control points.
The degree of a bezier curve is one less than the number of control points.
A set of bernstein polynomials are used for the basis functions of bezier curves.

A cubic bezier curve is defined using a sequence ($P_0$, $P_1$, $P_2$, $P_3$) of four control points.
Using the basis function, ${n \choose k}(1-u)^{k}u^{n-k}$ for $k$th control point $P_k$, $k \in \mathbb{Z} \cap [0,n]$, the parametric equation of the curve, on parameter $0 \le u  \le 1 $, is of the form, $b(u)$=$(1-u)^{3}P_0$ + 3$(1-u)^{2}uP_1$ + $3(1-u)u^{2}P_2$ + $u^{3}P_3$. In matrix form, this is expressed as, 
\begin{equation}
b(u) = \begin{pmatrix}
u^3 & u^2 & u & 1 
\end{pmatrix}
\begin{pmatrix}
1 & -3 & 3 & -1 \\
0 & 3 & -6 & 3 \\
0  & 0  & 3 & -3  \\
0 & 0 & 0 & 1 
\end{pmatrix}
\begin{pmatrix}
P_0 \\
P_1 \\
P_2  \\
P_3 
\end{pmatrix}
\end{equation}

where, matrix M=$\begin{pmatrix}
1 & -3 & 3 & -1 \\
0 & 3 & -6 & 3 \\
0  & 0  & 3 & -3  \\
0 & 0 & 0 & 1 
\end{pmatrix}$, corresponds to the geometry matrix (GM) of the curve.

Based on Equation 2, we have $b(0)$=$P_0$, and $b(1)$=$P_3$, or the control points at the beginning and end, are points on the curve.
Further, the tangents at the end of the parameter values, or the first derivatives at respective values, that could be seen to be, are $b^{'}(0)$=$3(P_1-P_0)$, and $b^{'}(1)$=$3(P_3-P_2)$.
A piecewise cubic curve of $C^1$ continuity at the junctions, can be constructed using cubic bezier curve segments, by ensuring equality of the respective tangents, similar to that of cardinal splines.
Thus, for any piecewise cubic curve, a bezier curve segment, with tuple of control points ($P_{j}$, $P_{j+1}$, $P_{j+2}$, $P_{j+3}$), to approximate a cardinal curve segment, from corresponding control point $p_{k-1}$ to $p_k$, requires $P_{j}$=$p_{k-1}$, $P_{j+3}$=$p_{k}$,  $P_{j+1}$= $P_{j}$ + $\frac{\tau}{3}$($p_{k+1}$ - $p_{k-2}$), and $P_{j+2}$= $P_{j+3}$ - $\frac{\tau}{3}$($p_{k+1}$ - $p_{k-1}$).   

\subsection{B-spline Curve}
B-spline curves use a geometric representation to approximate curves of specific degree, composed of a fixed number of curve segments of that degree in a given order.
B-spline curves identical to piecewise cubic bezier curves are possible to be constructed using the control points of the bezier curves representing the corresponding segments in the respective piecewise curves.
The representation using a cubic b-spline curve, for a curve identical in shape of $C^{1}$ continuous piecewise cubic bezier curve, improves the level of continuity of the composite curve.
The representation obtained using b-spline curve becomes $C^2$ continuous, or continuity upto curvatures, is raised over the existing continuity at the level of slopes.

For, a $C^1$ continuous piecewise bezier curve with sequence ($P_0$, $P_1$, ..., $P_n$) of $n$+1 control points, such that a sequence ($P_j$, $P_{j+1}$, $P_{j+2}$, $P_{j+3}$) of every four control points, subsequent to every fourth control point from the start index of the previous sequence, represents a piece which is a cubic bezier curve,  i.e, $j$ is of the form $j$=$3p$, where $p \in \mathbb{Z} \cap [0,n/3-1]$, can also represent the sequence of control points for a cubic b-spline curve.
The b-spline curve of cubic degree, with sequence ($P_0$, $P_1$, ..., $P_n$) of the control points, with order $k$=3+1=4, then describes ($n$-$k$+2), $C^2$ continuous curve segments, where each curve segment is computed with b-spline basis functions, for every four control points, subsequent to every second control point from the start index of the previous sequence and in the sequence.
The parameter range for the $l$th segment, is specified, as $l$-1 $\le u \le l$.
The recursively defined basis functions for b-spline curve of the form, $n(u)$ = $\sum_{i = 0}^{n}{N_{i,k}(u)P_i}$, with  $0 \le u \le$ ($n$-$k$+2), based on Carl de Boor's ~\cite{Farin,Piegl} expression, is as the follows,  

\begin{equation}
\begin{split}
N_{i,k}(u) = & \frac{(u-t_i)N_{i,k-1}(u)}{t_{i+k-1}-t_i} + \frac{(t_{i+k}-u)N_{i+1,k-1}(u)}{t_{i+k}-t_{i+1}}, where \\
&  N_{i,k}(u) = 1, t_i \le u \le t_{i+1}, and \\
& N_{i,k}(u) = 0, otherwise
\end{split}
\end{equation}

where, 
\begin{itemize}
\item $t_i$ (0 $\le$ $i$ $\le$ $n+k$) are called knot values
\item $t_i$ = 0 if $i < k$;
\item $t_i$ = $i-k+1$ if  $k \le i \le n$;
\item $t_i$ = $n-k+2$ if  $i > n$;
\end{itemize}

\subsection{Dominant Point}
A dominant point is an element of a set of the given data points, that is considered to dominate the shape characteristic of a b-spline curve approximating the corresponding set of data points ~\cite{Park}. A dominant point set contains dominant points, and is a subset of the given set of all data points, such that the b-spline curve approximating the subset of the points has an error within a given bound, compared to the curve approximating the given set of all data points. The complement of a dominant point set, is a set obtained by subtraction of the dominant point set from the set of all data points.

\subsection{Square Error for Dominant Point}
The error due to approximation of a b-spline curve using a set of dominant point could be computed as square error corresponding to the complement of the dominant point set. The corresponding points, for the elements in the complement of the dominant point set, on the approximating curve could be computed, based on respective parameter values in the piecewise curves. The elements in the dominant point set, could be arranged in a list, based on the relative order of every pair of points, in the list of all given points. A list of indices mapping to elements in a complement of dominant point set, is possible, between a pair of indices corresponding to two consecutive dominant points. Such a list is of length zero to any positive number, and square error for dominant points is produced only by every such list. The error due to an element of such a list, is computed about a point with appropriate parameter on the piecewise cubic bezier curve between the consecutive and corresponding dominant points.  The parameter value of the corresponding point on the curve is computed, based on the projection of the point on straight line joining the respective dominant points. The corresponding point is computed from the parameter value evaluated as the ratio of the projected length of the line joining the point with the dominant point representing the source of the parameter of the piecewise curve, on the line joining the dominant points, to the length of the line joining the consecutive dominant points.

\subsection{Turn Angle}
Turn angle for three consecutive data points in $\mathbb{R}^{2}$ is produced by the angle, by which line joining the first two points are to be rotated about the second point, from the straight line joining them, to reach to the third point. Turn angle $\theta_{j}$, at $j$th vertex, is computed using the following expression:
\begin{equation}
    \theta_{j}=\pi - cos^{-1}\frac{(\vec{FS} \cdot \vec{TS})}{(|\vec{FS}| |\vec{TS}|)}
\end{equation}
where,
\begin{itemize}
\item $\vec{v_j}$: Represents coordinate of $j$th point, as an ordered pair ($p_j$,$q_j$), of independent coordinate $p_j$, and dependent coordinate $q_j$.
\item $\vec{FS}$: $\vec{v_{j-1}}$ - $\vec{v_{j}}$
\item $\vec{TS}$: $\vec{v_{j+1}}$ - $\vec{v_{j}}$
\end{itemize}

The turn angle for point $P_2$, is the exterior angle of $P_1P_2P_3$ corresponding to the triangle $P_1P_2P_3$. 

\subsection{Selection and Partition of Dominant Point}

Dominant point set of a given cardinality are selected by minimizing square error for complement of dominant point set. For a dominant point set of size $m$, from a given point set of size $n$, the square error, $e_m$ is expressed as, 
\begin{equation}
e_m = \sum_{i=0, i'=next(L,i), j=next(L,i'), i=i'}^{j=n}\sum_{k=1}^{j-i-1}((x_{i+k,d}-x_{i+k})^2+(y_{i+k,d}-y_{i+k})^2)
\end{equation}

where,
\begin{itemize}
    \item $L$ : It is a list of indices corresponding to the all points, mapping to the dominant points.
    \item $(x_{i+k}, y_{i+k})$ : It is the $(i+k)$th point in the list of input data points, which is not contained in corresponding list $L$ of dominant points, and is relevant in between the elements $i$ and $j$ in $L$.
    \item $(x_{i+k,d}, y_{i+k,d})$ : It is a mapping of the point $(x_{i+k}, y_{i+k})$, on the piecewise curve between dominant points $i$ and $j$.
    \item $next(L,i)$ : It returns the element next of $i$ in list of indices $L$ of dominant points.
\end{itemize}

In order, to minimize the square error, due to $m$ dominant points, we modify an initial guess, to reach to local minima. We use the following conditions in order of non-increasing priority for a good guess. We use associative array $arr$ of turn angles and indices of all data points, sorted in descending order on turn angle. 

\begin{itemize}
\item Select first $m_1$ points in the sorted order from array $arr$, as primary dominant points.
\item Select $m_2$ points with at least one previous and one next vertices of each of the primary dominant points, from order of given input data points, as support dominant points. 
\item Select remaining ($m$-$m_1$-$m_2$) points, from array $arr$, in the sorted order, subsequent to index $m_1$ in the array, as secondary dominant points.    
\end{itemize}

We, minimize the error $e_m$, due to $m$ dominant points, from an initial guess, by selection and deselection of additional points, until any minimal error. Towards the objective to minimize error, we select a dominant point from a pair of dominant points, that give rise to maximal error among all pairs, and deselect a dominant point from a pair of dominant points with minimal error, and recompute the new error. We update the dominant points and the corresponding error of $e_m$ to $e_{m}^{'}$, if $e_{m}^{'}$ is lower than $e_{m}$. We stop at selection of dominant points at the step when no update yields any improvement of the error.

Selection of dominant points from input data points in Figure \ref{dpntfig}.a, is shown in Figure \ref{dpntfig}.b.
Point indices $i$-1, and $i$+3, correspond to primary dominant points, based on ordering of turn angles.
Point indices $i$, and $i$-2, and point index $i$+4, correspond to support dominant points for points with indices $i$-1, and $i$+3, respectively.
Square error is included for points with indices $i$+1, and $i$+2, corresponding to computation using the points.

\subsection{Corresponding Coordinate Planes}
Corresponding coordinate planes are produced by two independent $\mathbb{R}^{2}$ systems, representing affine maps to respective planar curves, from curves in $\mathbb{R}^{3}$.
Corresponding coordinate planes of an $\mathbb{R}^{3}$ system, is a pair of coordinate planes, using two distinct pairs of its coordinate axis, with any common coordinate axis, as the independent coordinate, in respective $\mathbb{R}^{2}$ systems. A pair of corresponding coordinate planes of an $\mathbb{R}^{3}$ system, with coordinate axes $x_i$, $x_j$, and $x_k$, are $\mathbb{R}^{2}$ systems of coordinate axes $x_j$ with $x_i$, and coordinate axes $x_j$ with $x_k$.

\subsection{Merge rule for Control Points in lower dimension to higher dimension}
We merge coordinates of control points of piecewise bezier curve segments in corresponding coordinate planes, to obtain coordinates of control points for the respective curve segments in  $\mathbb{R}^{3}$.   The curve in the $\mathbb{R}^{3}$ system, or in the higher dimension, is represented using the control points of the curve in the corresponding system.  

For tuple ($P_{i}^{a}$, $P_{i+1}^{a}$, $P_{i+2}^{a}$, $P_{i+3}^{a}$) of control points in coordinate plane $a$ and axes $x_k$ with $x_j$, and tuple ($P_{i}^{b}$, $P_{i+1}^{b}$, $P_{i+2}^{b}$, $P_{i+3}^{b}$) of control points in coordinate plane $b$ and axes $x_k$ with $x_l$, the following steps are considered:

\begin{enumerate}
\item Compute $p$ equals $\frac{e+f}{2}$ where $e$ is difference in $x_k$ coordinate of point $P_{i+1}^{a}$ and $P_{i}^{a}$, $f$ is difference in $x_k$ coordinate of point $P_{i+1}^{b}$ and $P_{i}^{b}$, and $q$ equals $\frac{g+h}{2}$ where $g$ is difference in $x_k$ coordinate of point $P_{i+2}^{a}$ and $P_{i+3}^{a}$, $h$ is difference in $x_k$ coordinate of point $P_{i+2}^{b}$ and $P_{i+3}^{b}$.  Thus, $p$ is the signed magnitude of arithmetic mean of the intervals in the independent coordinates of the ($i$+1)th point with the $i$th point in the coordinate planes $a$ and $b$, and $q$ is the signed magnitude of arithmetic mean of the intervals in the independent coordinates of the ($i$+2)th point with the ($i$+3)th point in the coordinate planes $a$ and $b$.
\item Generate new point $P_{i+1,p}^{a}$ with $x_k$ coordinate equals sum of $p$ and $x_k$ coordinate of point $P_{i}^{a}$, on line joining points $P_{i}^{a}$ and $P_{i+1}^{a}$, and new point $P_{i+2,q}^{a}$ with $x_k$ coordinate equals sum of $q$ and $x_k$ coordinate of point $P_{i+3}^{a}$, on line joining points $P_{i+3}^{a}$ and $P_{i+2}^{a}$.
\item Generate new point $P_{i+1,p}^{b}$ with $x_k$ coordinate equals sum of $p$ and $x_k$ coordinate of point $P_{i}^{b}$, on line joining points $P_{i}^{b}$ and $P_{i+1}^{b}$, and new point $P_{i+2,q}^{b}$ with $x_k$ coordinate equals sum of $q$ and $x_k$ coordinate of point $P_{i+3}^{b}$, on line joining points $P_{i+3}^{b}$ and $P_{i+2}^{b}$.
\item Collect $x_j$ and $x_k$, coordinates of point $P_{i}^{a}$, and $x_l$ coordinate of point $P_{i}^{b}$, to obtain coordinate of higher dimensional point $P_{i}^{a,b}$, and collect $x_j$ and $x_k$, coordinates of point $P_{i+3}^{a}$, and $x_l$ coordinate of point $P_{i+3}^{b}$, to obtain coordinate of higher dimensional point $P_{i+3}^{a,b}$.
\item Collect $x_j$ and $x_k$, coordinates of point $P_{i+1,p}^{a}$, and $x_l$ coordinate of point $P_{i+1,p}^{b}$, to obtain coordinate of higher dimensional point $P_{i+1,p}^{a,b}$, and collect $x_j$ and $x_k$, coordinates of $P_{i+2,p}^{a}$, and $x_l$ coordinate of $P_{i+2,p}^{b}$, to obtain coordinate of higher dimensional point $P_{i+2,p}^{a,b}$.
\item The tuple ($P_{i}^{a,b}$, $P_{i+1,p}^{a,b}$, $P_{i+2,q}^{a,b}$, $P_{i+3}^{a,b}$), correspond to control points, for the higher dimensional cubic bezier curve.
\end{enumerate}

The merge operation is applied using the control points only. A point coordinate ($e$, $f$, $g$) of a control point in the three dimensional space, is obtained from coordinates ($f$, $e$), and ($f$, $g$), of control points in the corresponding coordinate planes.   The operation thus requires the independent coordinates for the control points in respective pairs, on the corresponding coordinate planes, to be equal. This holds automatically, for every first and fourth control point, of every cubic bezier curve piece, for the pair of piecewise bezier curves, on the corresponding coordinate planes.  The reason for this is, these points correspond to data points, and are points on the planar curves, due to the interpolation only. The second and the third control points for the curves are computed, using constraints on the tangent vectors at end points of the curve pieces and equality of tangents between successive pieces, at the data points, representing the junctions of the curve pieces. Equality for independent coordinates for the second and the third points does not hold, automatically, and their coordinates are required to be constrained to achieve equality. The independent coordinates of the second control points of the curves and the third control points of the curves on the pair of coordinate planes could be fixed to single values. They could be made to impact only the magnitudes of the tangents proportionately, with no change to the directions, at respective start and end points of the curve pieces in the piecewise curves, on the corresponding planes.

Fixing a value for independent coordinate of the second control point, at arithmetic mean of the intervals in the independent coordinates of the second point from the first point on the pair of coordinate planes, about the first control point, could be considered, in case of update in the second control point. Similarly, fixing a value for independent coordinate of the third control point, at arithmetic mean of the intervals in the independent coordinates of the third point from the fourth point on the pair of coordinate planes, about the fourth control point, could be considered, in case of update in the third control point.
This would share any impact on shape preserving property due to change in magnitudes in the tangents at the data points, based on interpolation using the underlying cardinal spline equally, among the curves on the corresponding coordinate planes. 
The first step of the merge operation, computes this signed magnitude of arithmetic mean, from the intervals in the independent coordinates, on the corresponding coordinate planes.
In order to preserve the direction of the tangents, at the corresponding first and fourth points of the curve pieces, identical as before, the dependent coordinate of the respective second and third control points, are computed, on the lines joining the first and second control point, and the third and the fourth control point, respectively.  The dependent coordinates are computed using the corresponding values of the independent coordinates, that is being fixed. The second and the third step of the merge operation, compute the direction preserving, second and third control points, using the independent coordinates computed at the first step of the merge operation, on the first and the second coordinate planes, respectively. The equality of the tangent values at last control point of a piece and first control point of the next piece, that is required to be satisfied, at the junction point of the curve pieces for $C^1$ continuity, is automatically satisfied. In this case, it becomes possible, due to selection of arithmetic mean for computing the independent coordinates, from the coordinates of the control points on the respective coordinate planes. The fourth and fifth step of the merge operation, copies the respective coordinates, of the first and second coordinate planes respectively, into the coordinates of the three dimensional control points. The control points of cubic bezier curve piece in the three dimensional space, is obtained by merging of control points of the curve of the corresponding coordinate planes at the end of the sixth step.

Merging of control points of curve in $\mathbb{R}^{4}$ system with coordinate axes ($x_j$, $x_k$, $x_l$, $x_m$), is possible by merging curves from corresponding planes $a$, $b$, and $c$, or $\mathbb{R}^{2}$ systems with coordinate axes ($x_j$, $x_k$), ($x_j$, $x_l$), and ($x_j$, $x_m$), respectively. Using the algorithm to merge control points of corresponding coordinate planes, discussed in this section, merging is required to be computed, for the planes $a$ with $b$, and $a$ with $c$, separately, in this case.  However, for target system $\mathbb{R}^{4}$, in contrast to $\mathbb{R}^{3}$ of before, the independent coordinates for the second, and the third control points, are required to be fixed, using arithmetic mean, of three terms, arising from the three planes $a$, $b$, and $c$, in contrast to two terms of the earlier from $a$, and $b$ planes only. This change corresponds only to the first step of the algorithm, and then steps second to fifth, are applied independently, for the $\mathbb{R}^{3}$  systems, resulting coordinate axes ($x_j$, $x_k$, $x_l$) from the coordinate planes $a$ with $b$,  and coordinate axes ($x_j$, $x_k$, $x_m$) from the coordinate planes $a$ with $c$. Copy of the control point coordinates resulting from the two $\mathbb{R}^{3}$ systems, would be output for merged control point coordinates in the $\mathbb{R}^{4}$ system, at the end of the sixth step.  By induction, therefore, merging of control points could be computed for a system represented by $\mathbb{R}^{n}$ with coordinate axes ($x_1$, $x_2$, ..., $x_n$), and control points of curves from ($n$-1) coordinates planes as ($x_j$,$x_1$), ($x_j$,$x_2$), ..., ($x_j$,$x_{j-1}$), ($x_j$,$x_{j+1}$), ..., ($x_j$,$x_{n}$) using any common independent coordinate axis, $x_j$, $j \in \mathbb{N} \cap [1,n]$. The independent coordinates of the second and the third control points of the curve pieces, in that case, are required to be computed using arithmetic mean of ($n$-1) terms, corresponding to the ($n$-1) coordinate planes.

\section{Present Work}
\label{sec2_}

In this Section, we present our algorithm, to fit cubic B-spline curve, to a chain of data points in three, or higher dimensional space, using terminologies discussed in Section 2. At first we present steps to construct cubic b-spline curves, to the data points in corresponding coordinate planes. Subsequently, we present steps to compute the control points and the curve in the higher dimensional space, from curves in the corresponding coordinate planes, using the merge rule that we have discussed before.

 
\subsection{Algorithm for fitting curve to data points in a coordinate plane (FC2)} 
The steps in fitting a cubic b-spline curve to data points in a coordinate plane are as follows:
\begin{enumerate}
\item Fit cardinal spline interpolating the data points with default value of 0.5 for constant representing $tension$ parameter of the curve.
\item Use coordinates and tangents at the data points, to fit tangent continuous cubic bezier curve pieces to every pair of consecutive data points.   
\item Use coordinates of control points of the piecewise cubic bezier curve, to fit a curvature continuous cubic b-spline curve.
\end{enumerate} 
 
\subsection{Algorithm for fitting curve to data points in the three dimensional space (FC3)} 

The steps in fitting a cubic b-spline curve to data points in the three dimensional space is as follows:
\begin{enumerate}
\item Select data points representing corresponding coordinate planes to compute cubic b-spline curves in respective planes using steps specified in FC2 algorithm.
\item Collect coordinates of control points of piecewise cubic bezier curves on respective coordinate planes.
\item Merge coordinates of control points of cubic bezier curves at corresponding coordinate planes, to compute control points for the space curve.      
\item Use coordinates of control points in three dimensional space of the piecewise cubic bezier curve, to fit a curvature continuous cubic b-spline curve.
\end{enumerate}

\section{Related Work}
Park et. al. ~\cite{Park} proposed DOM scheme, to improve accuracy and quality of B-spline curve fitting, for planar curves. The approach minimizes square error, by subset of points, named dominant points, selected on the curve, for parameterization. More points are selected from regions of higher curvature, to increase the information content, related to the shape of the curve. Chord length, and centripetal parameterization is used, for the points, indexed from an array of given points, in the fitting. Approach for fitting smooth B-spline curves/surfaces, to unordered/unstructured cloud of points, was proposed by Ma et. al. ~\cite{Ma}. Parameterization was considered with reference to point coordinates projected onto a base curve/surface, from the input space. Ravari et. al. ~\cite{Ravari} minimized cost, of the least squares fit, based on a measure of fitness for the sample, corresponding to information reduced of salient points.  

Sung et. al. ~\cite{Sung} presented algorithms to enhance performance of optimization of sum of squared distances, for nonlinear problem of geometric fitting of implicit surfaces and planar curves, using Gauss-Newton method.  Form, position, and rotation-based parameters are considered, in Lagrange multipliers, with coordinate-based and distance-based error measures. 

Yang et. al. ~\cite{Yang} proposed approach for fitting, an active B-spline curve/contour, to a target. The approach does not require parameterization, and inserts/deletes control points, and knots, to converge to the target curve. The approach improves speed of convergence, of square distance minimization (SDM) technique of Pottmann et. al. ~\cite{Pottmann}. Fast marching method ~\cite{Sethian}, is used on distance field computed among curves, from localized footpoint.

Juttler et. al. ~\cite{Juttler} proposed approaches for implicit description of curves and surfaces, based on solution of simple constraints, and linear equations.   The approach uses shape constraints, in contrast to explicit parameterization for surfaces at first. The approach emphasized computation based on sign generating real function, and considerations for simple algorithms using intersection with straight lines. Regions favoring simple computation are grown incrementally, based on processing of scanlines.

Pateloup et. al. ~\cite{Pateloup} used B-spline curves, for interpolation of $C^{2}$ continuous tool paths required in pocket machining. Numerical controlled paths are produced based on single curves, for milling operation with required accuracy, and in minimum machining time.

\section{Experimental Results}
In this Section, we present results of experimentation using our algorithm for fitting cubic b-spline curve to data points in the three dimensional space. We also present results of experimentation using our algorithm for selection and partitioning of dominant points, for approximating the fitting of cubic b-spline curves to given data points in coordinate planes, based on minimization of square error. 
We implemented our algorithms in matlab, and have run our programs on Octave 4.0.2 installed on standard Windows 7 laptop PC. Certain data sets for our experimentation on approximating b-spline curve using dominant points, were followed from related work of Park et. al. ~\cite{Park}, and Ma et. al. ~\cite{Ma}. 

\begin{figure}
    \centering
    \begin{subfigure}[b]{0.3\textwidth}
        \includegraphics[width=\textwidth]{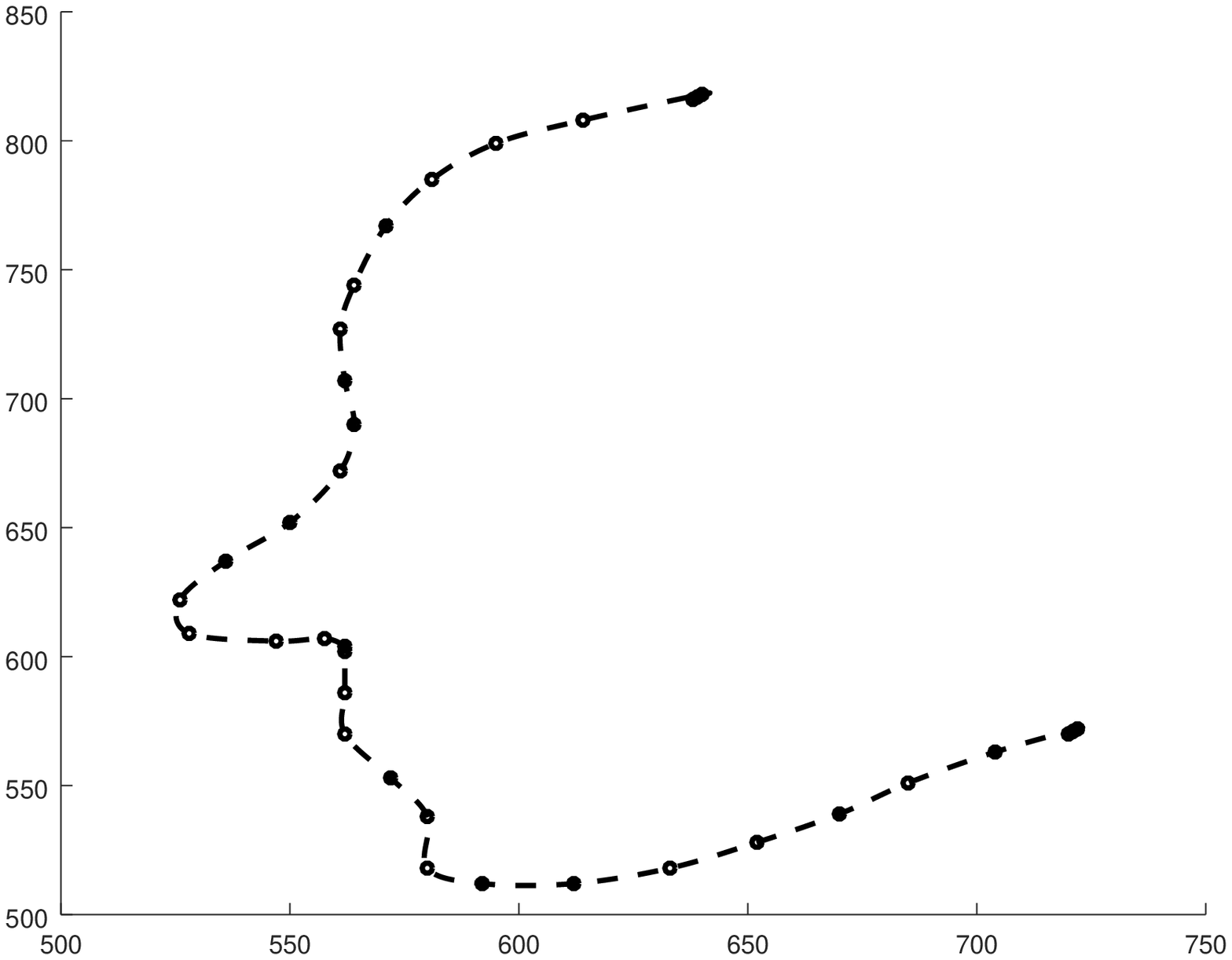}
        \caption{Interpolated Cardinal spline}
        \label{fig:gull}
    \end{subfigure}
    ~ 
    \begin{subfigure}[b]{0.3\textwidth}
        \includegraphics[width=\textwidth]{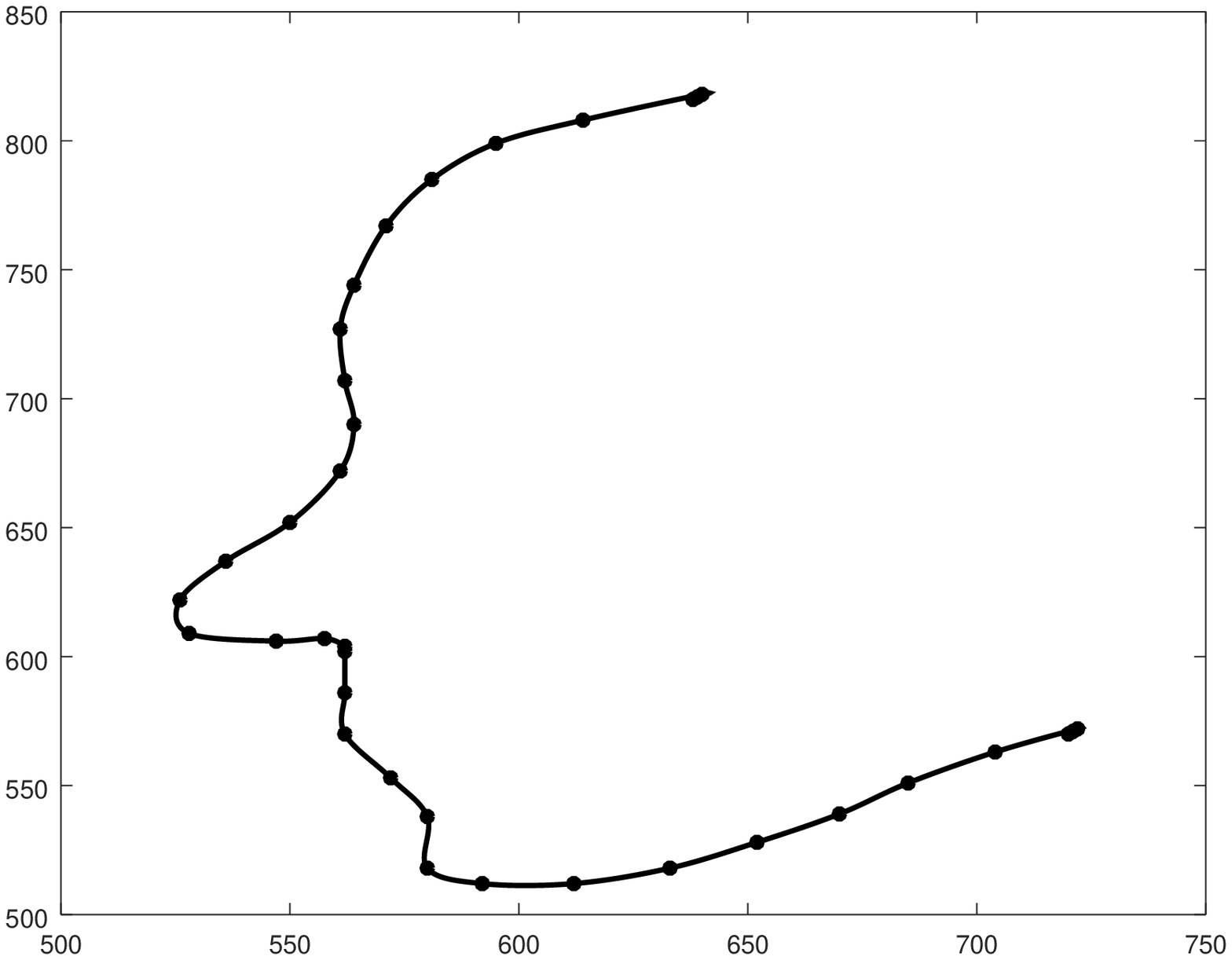}
        \caption{Piecewise cubic Bezier fit}
        \label{fig:tiger}
    \end{subfigure}
    ~ 
    \begin{subfigure}[b]{0.3\textwidth}
        \includegraphics[width=\textwidth]{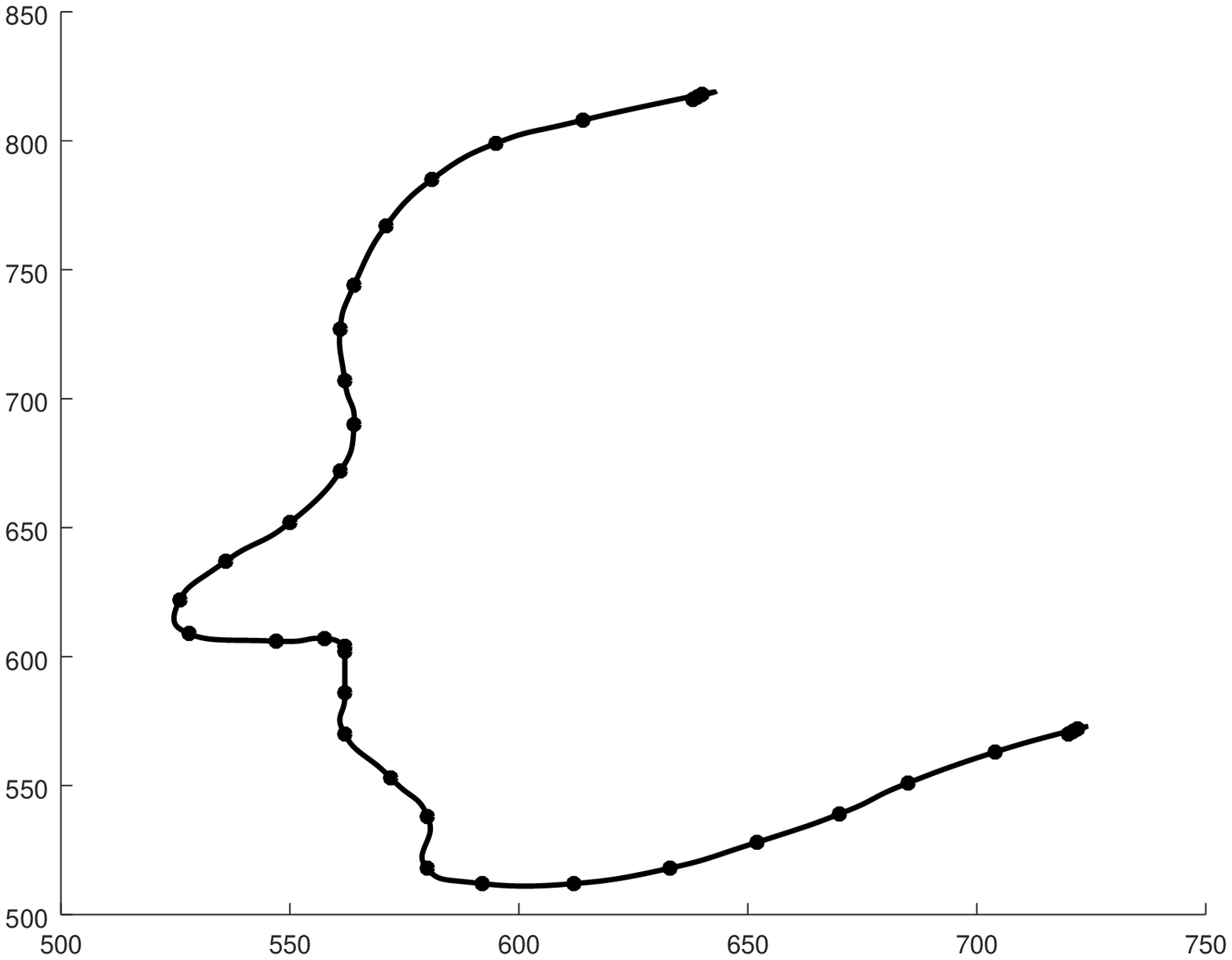}
        \caption{Cubic B-spline fit}
        \label{fig:mouse}
    \end{subfigure}
    \begin{subfigure}[b]{0.3\textwidth}
        \includegraphics[width=\textwidth]{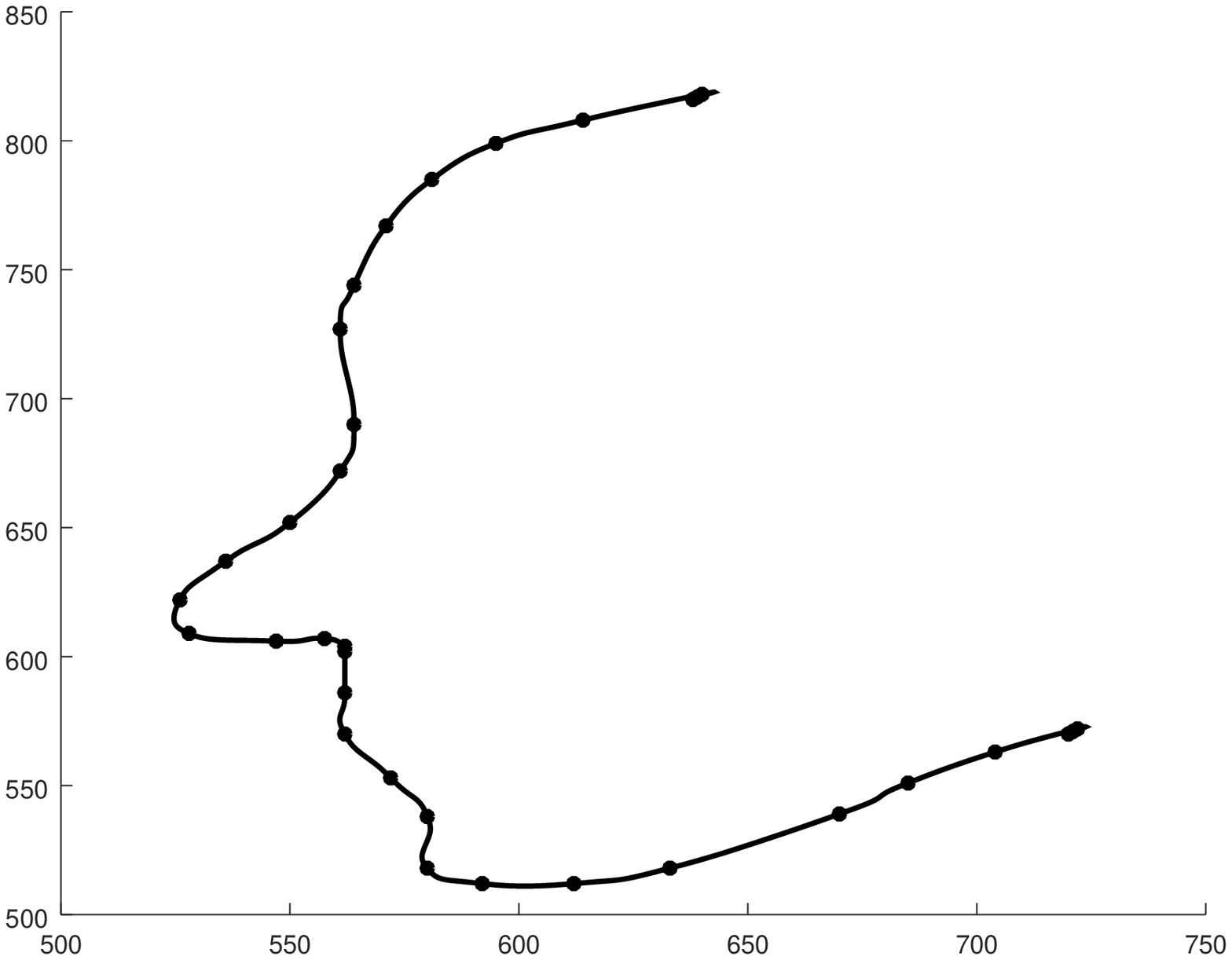}
        \caption{Approximated b-spline with 90\% points (error $\approx$ 1.0)}
        \label{fig:gull}
    \end{subfigure}
    ~ 
    \begin{subfigure}[b]{0.3\textwidth}
        \includegraphics[width=\textwidth]{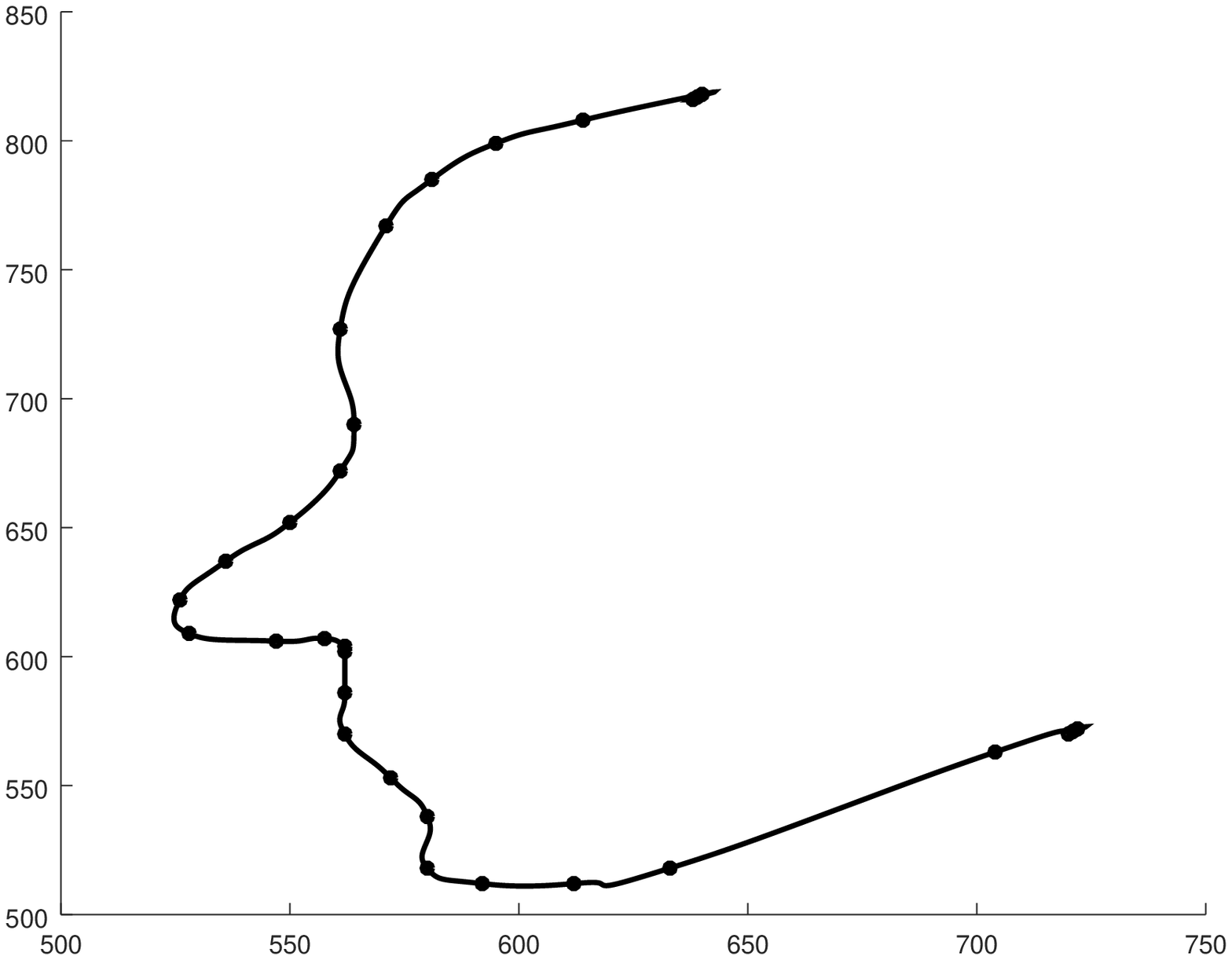}
        \caption{Approximated b-spline with 85\% points (error $\approx$ 9.0)}
        \label{fig:tiger}
    \end{subfigure}
    ~ 
    \begin{subfigure}[b]{0.3\textwidth}
        \includegraphics[width=\textwidth]{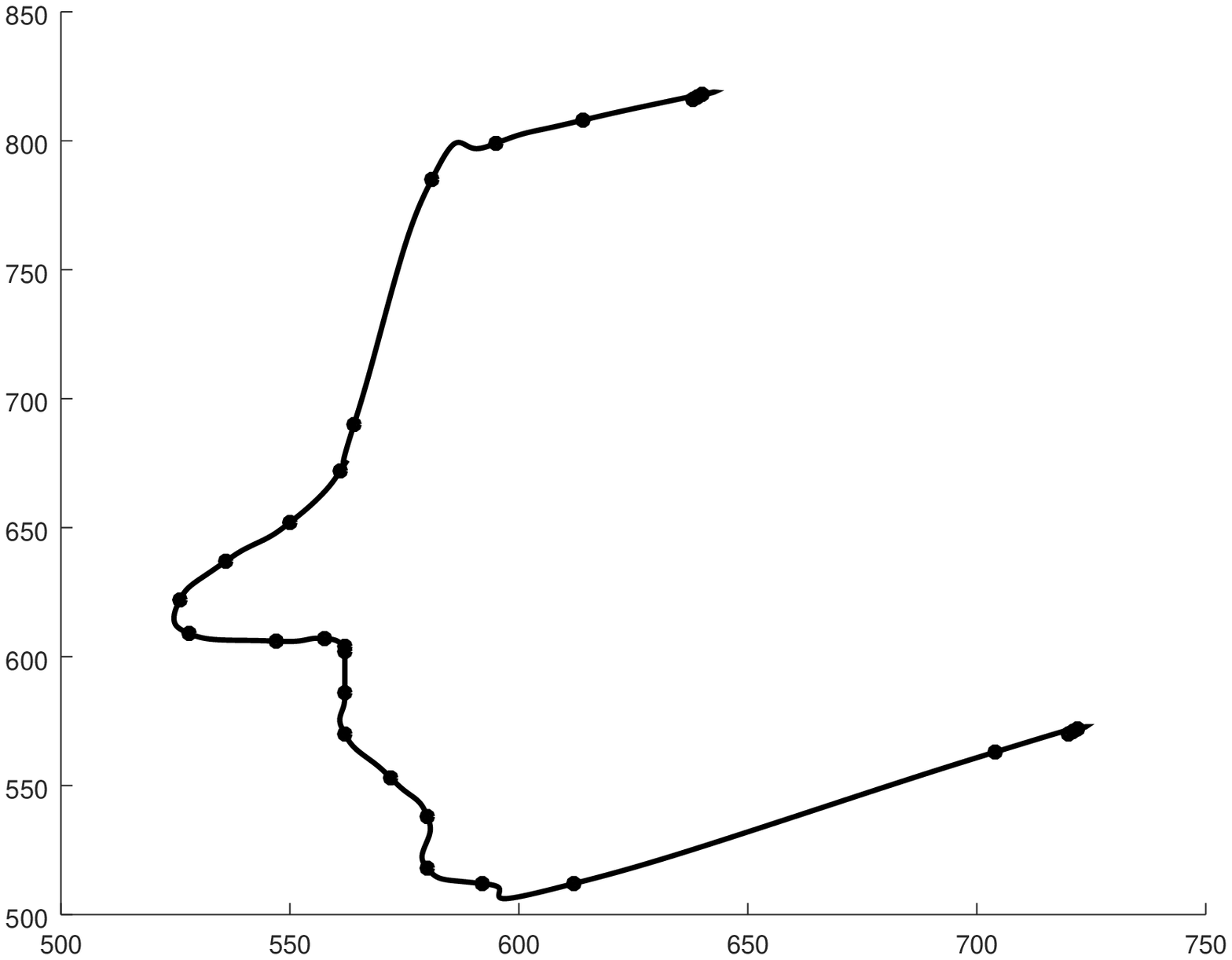}
        \caption{Approximated b-spline with 78\% points (error $\approx$ 339.6)}
        \label{fig:mouse}
    \end{subfigure}
    \caption{Experimentation 1}\label{res1_2dfig}
\end{figure}

\begin{figure}
    \centering
    \begin{subfigure}[b]{0.3\textwidth}
        \includegraphics[width=\textwidth]{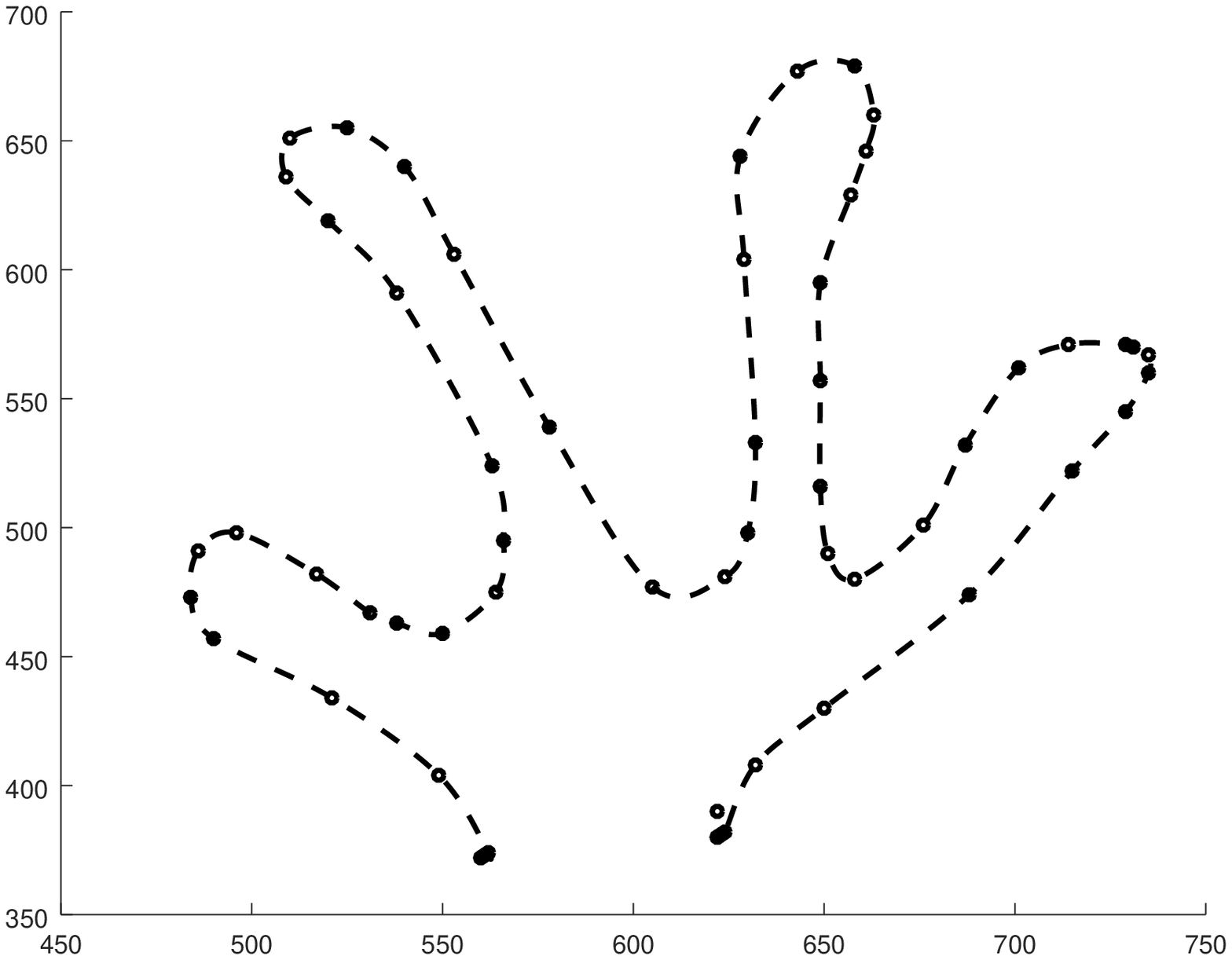}
        \caption{Interpolated Cardinal spline}
        \label{fig:gull}
    \end{subfigure}
    ~ 
    \begin{subfigure}[b]{0.3\textwidth}
        \includegraphics[width=\textwidth]{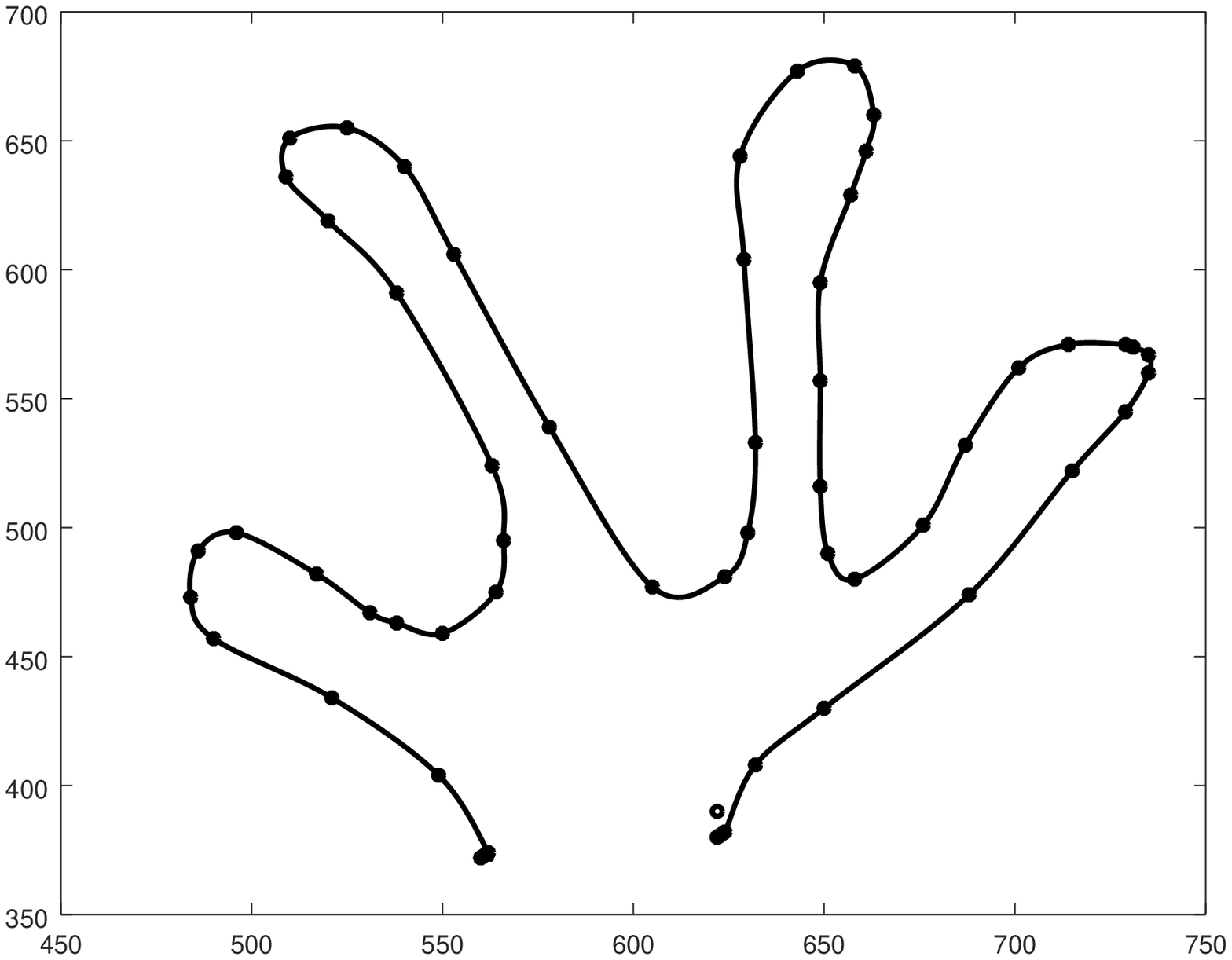}
        \caption{Piecewise cubic Bezier fit}
        \label{fig:tiger}
    \end{subfigure}
    ~ 
    \begin{subfigure}[b]{0.3\textwidth}
        \includegraphics[width=\textwidth]{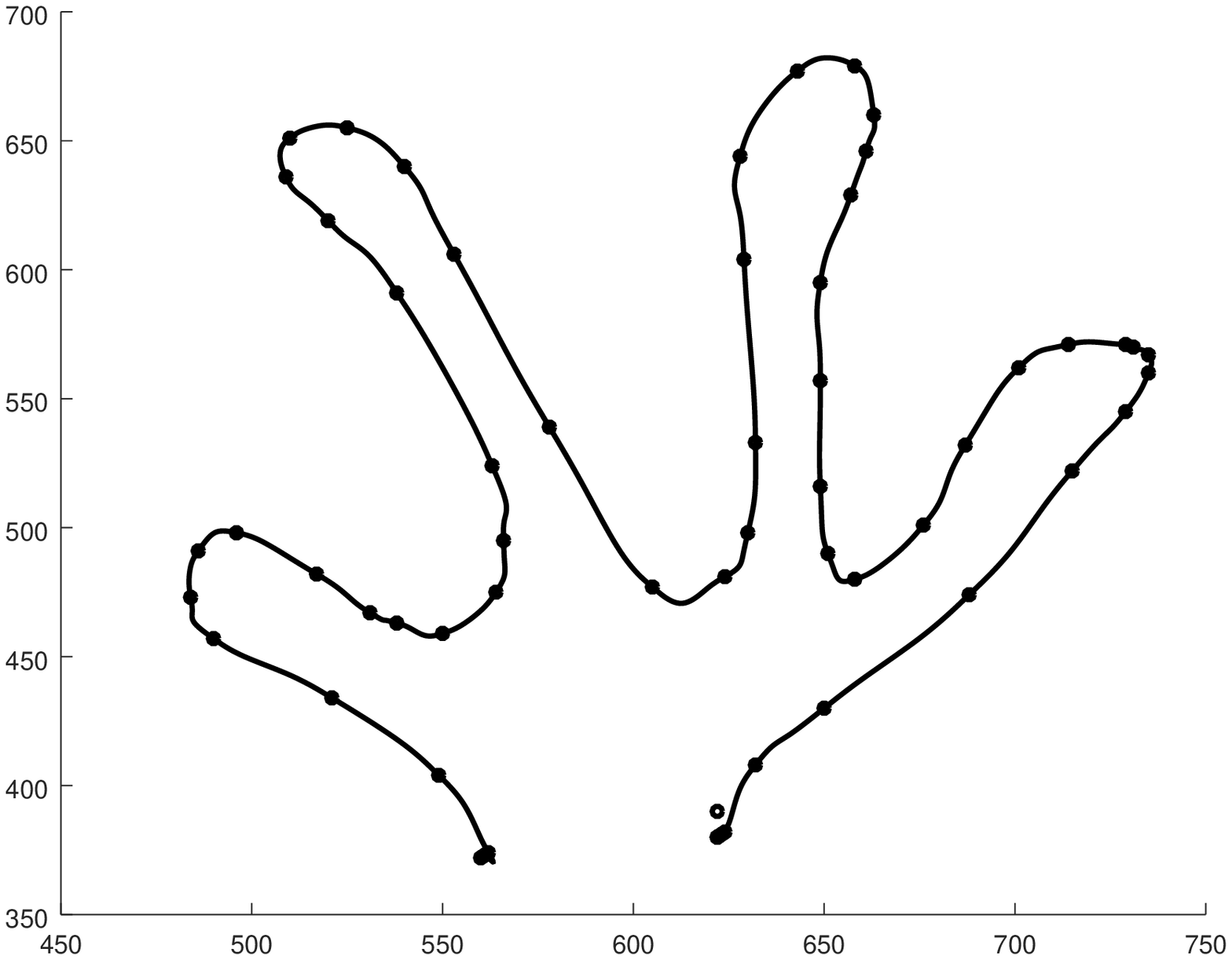}
        \caption{Cubic B-spline fit}
        \label{fig:mouse}
    \end{subfigure}
    \begin{subfigure}[b]{0.3\textwidth}
        \includegraphics[width=\textwidth]{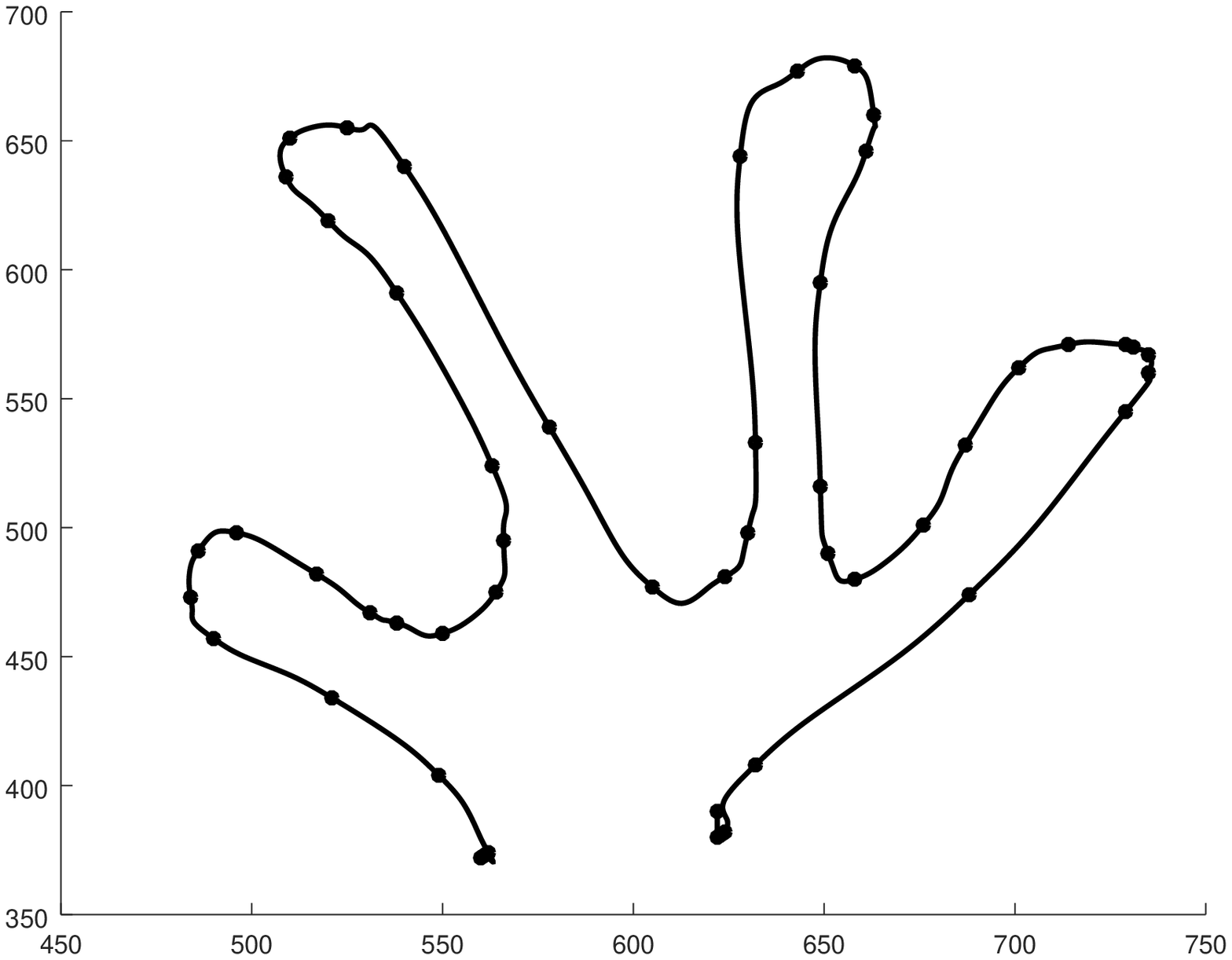}
        \caption{Approximated b-spline with 90\% points (error $\approx$ 0.9)
}
        \label{fig:gull}
    \end{subfigure}
    ~ 
    \begin{subfigure}[b]{0.3\textwidth}
        \includegraphics[width=\textwidth]{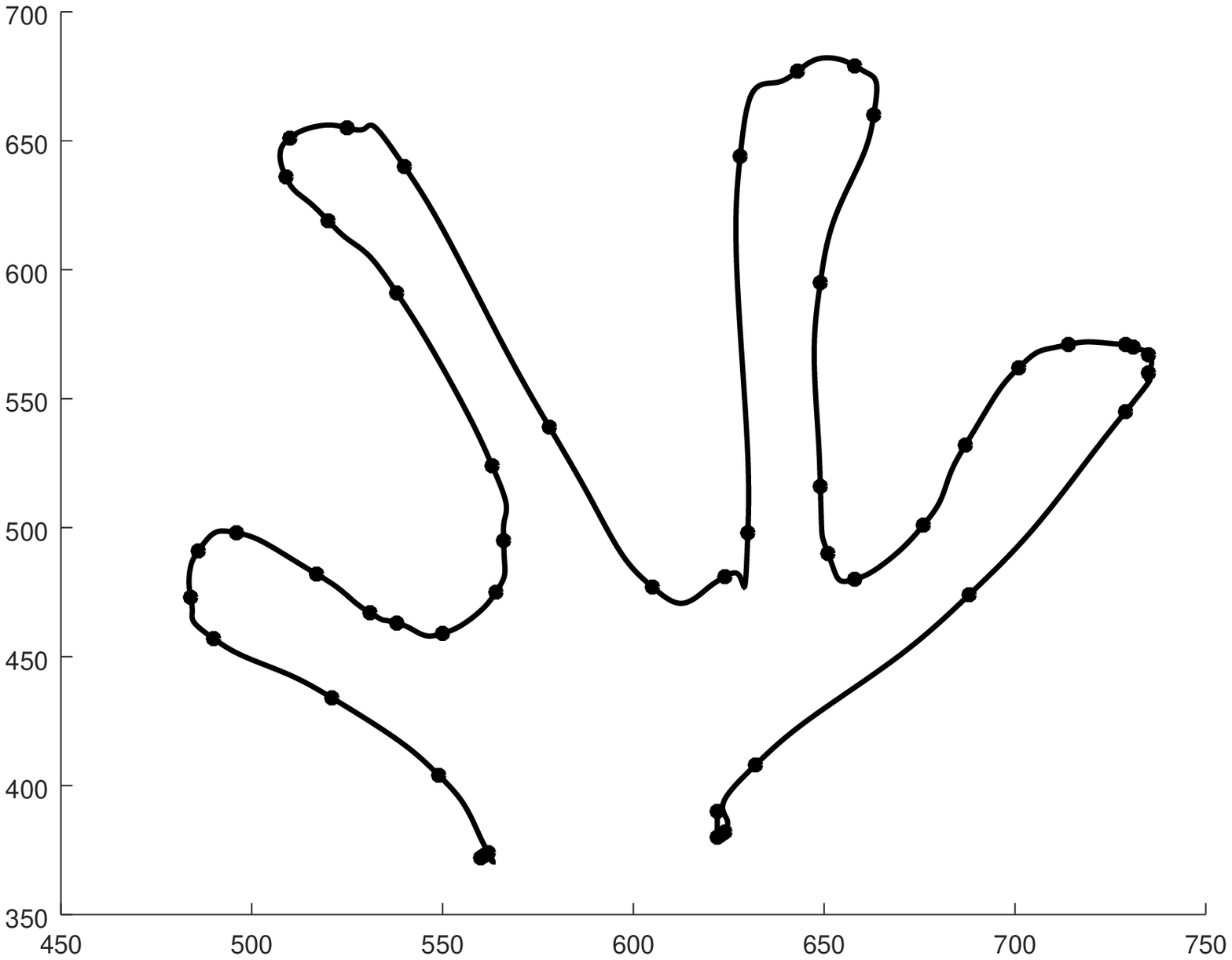}
        \caption{Approximated b-spline with 85\% points (error $\approx$ 8.4)
}
        \label{fig:tiger}
    \end{subfigure}
    ~ 
    \begin{subfigure}[b]{0.3\textwidth}
        \includegraphics[width=\textwidth]{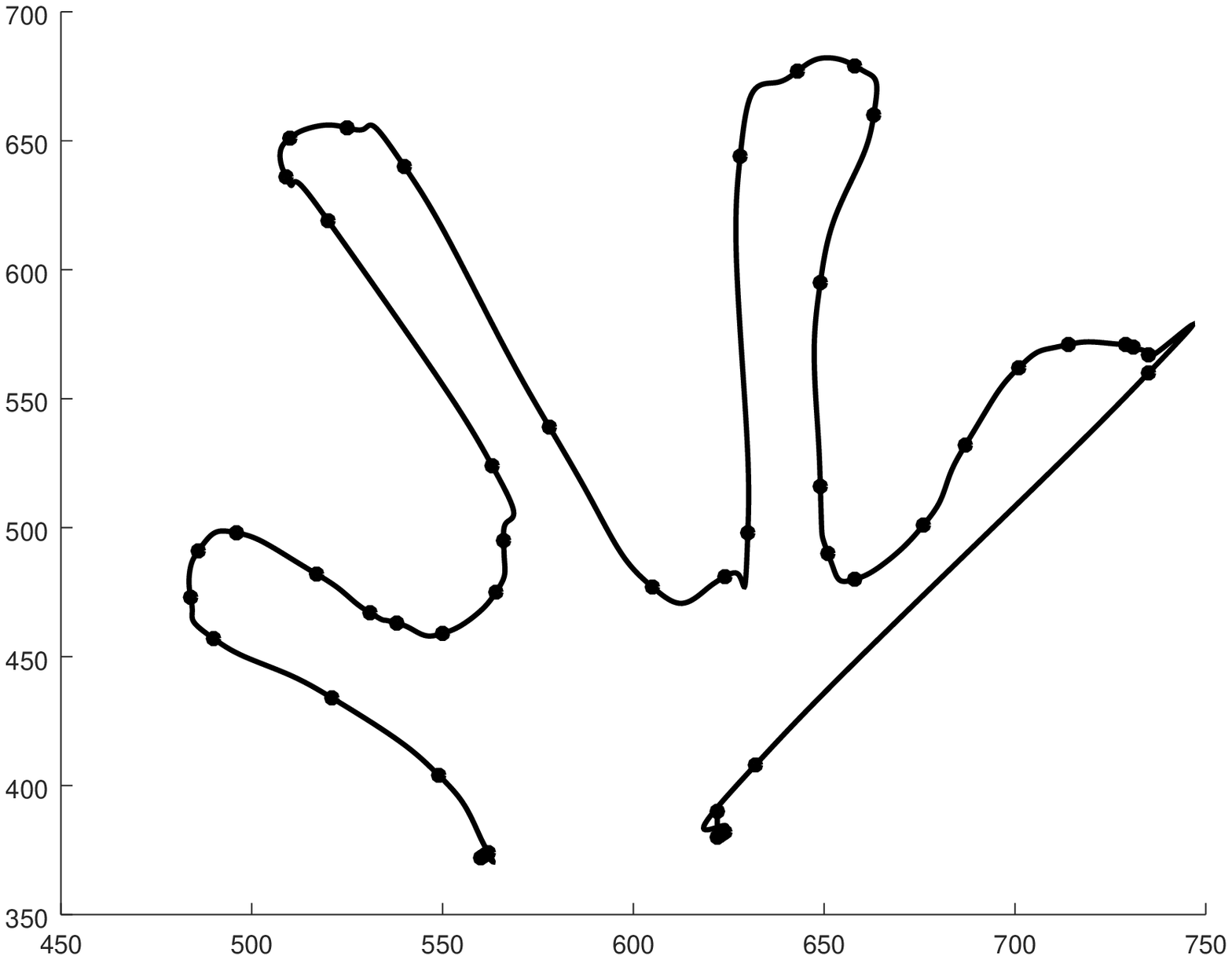}
        \caption{Approximated b-spline with 80\% points (error $\approx$ 161.3)}
        \label{fig:mouse}
    \end{subfigure}
    \caption{Experimentation 2}\label{res2_2dfig}
\end{figure}

\begin{figure}
    \centering
    \begin{subfigure}[b]{0.3\textwidth}
        \includegraphics[width=\textwidth]{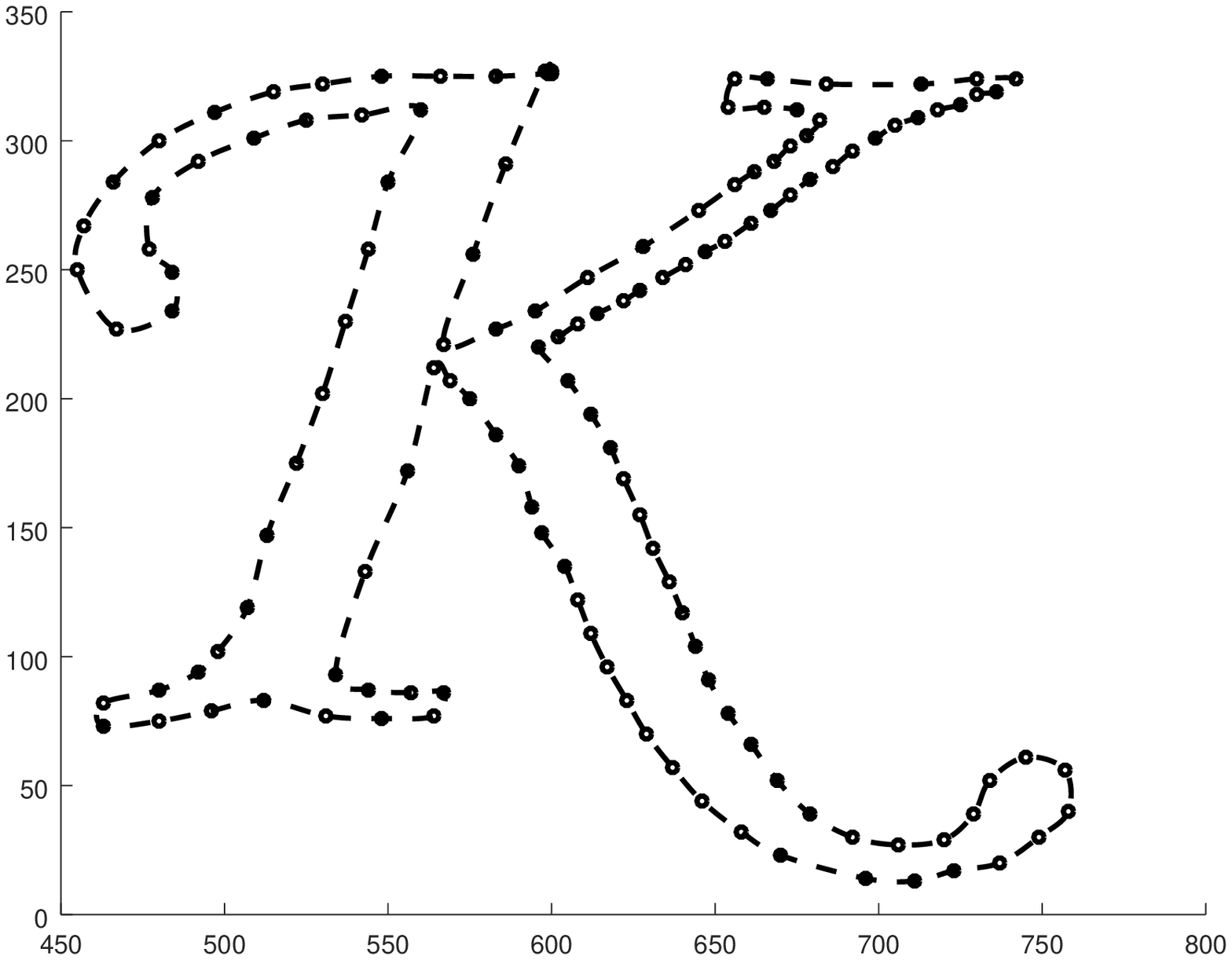}
        \caption{Interpolated Cardinal spline}
        \label{fig:gull}
    \end{subfigure}
    ~ 
    \begin{subfigure}[b]{0.3\textwidth}
        \includegraphics[width=\textwidth]{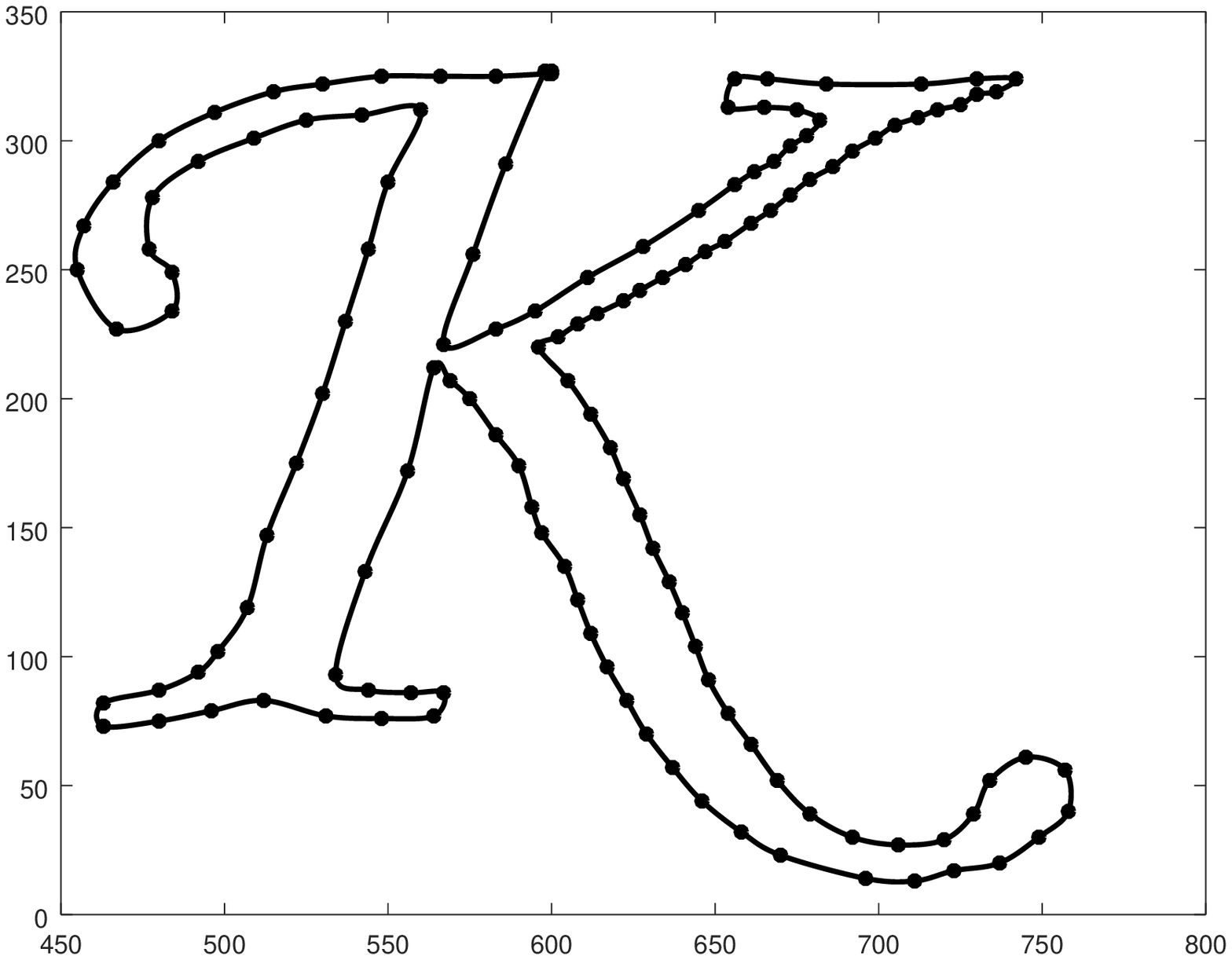}
        \caption{Piecewise cubic Bezier fit}
        \label{fig:tiger}
    \end{subfigure}
    ~ 
    \begin{subfigure}[b]{0.3\textwidth}
        \includegraphics[width=\textwidth]{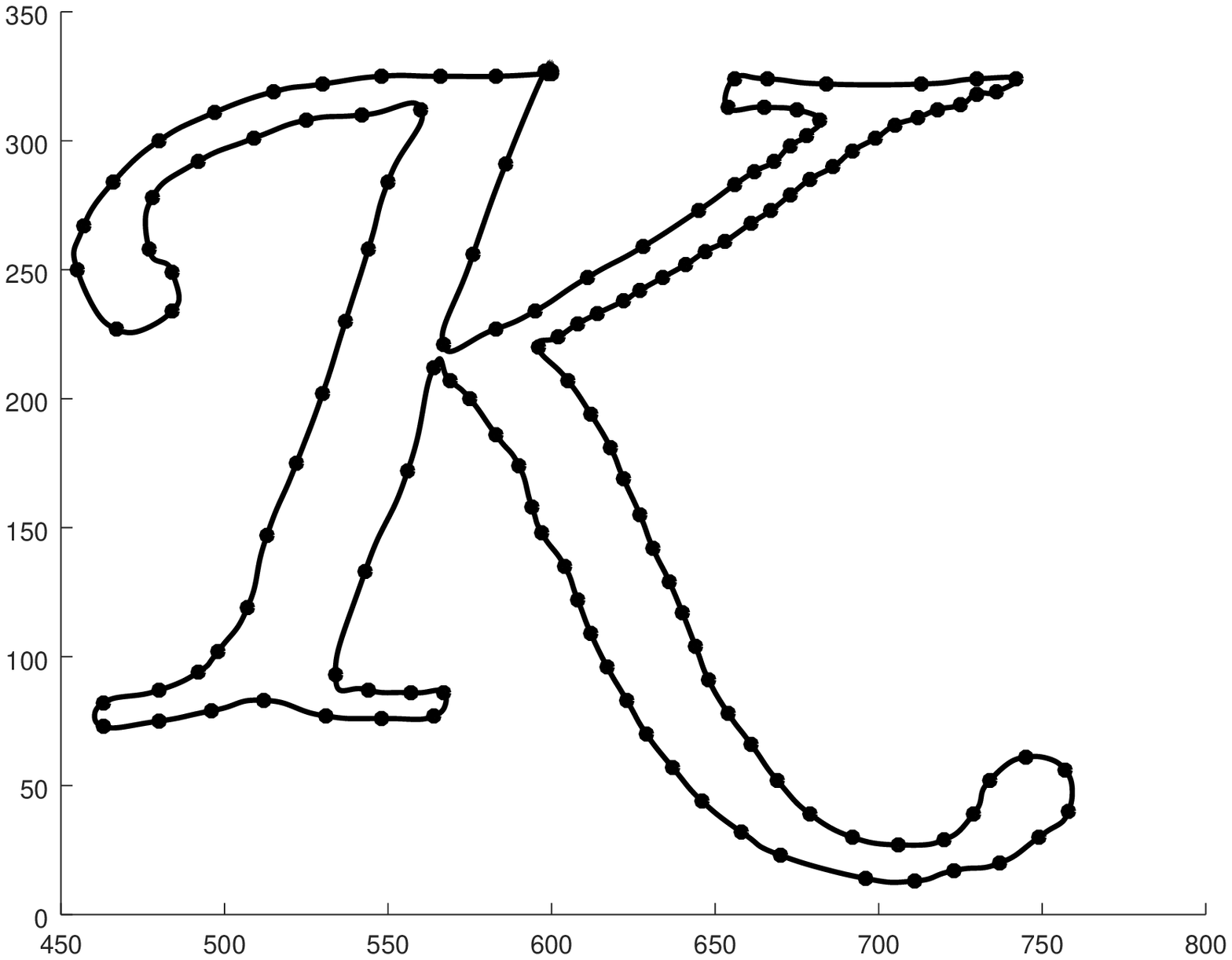}
        \caption{Cubic B-spline fit}
        \label{fig:mouse}
    \end{subfigure}
    \begin{subfigure}[b]{0.3\textwidth}
        \includegraphics[width=\textwidth]{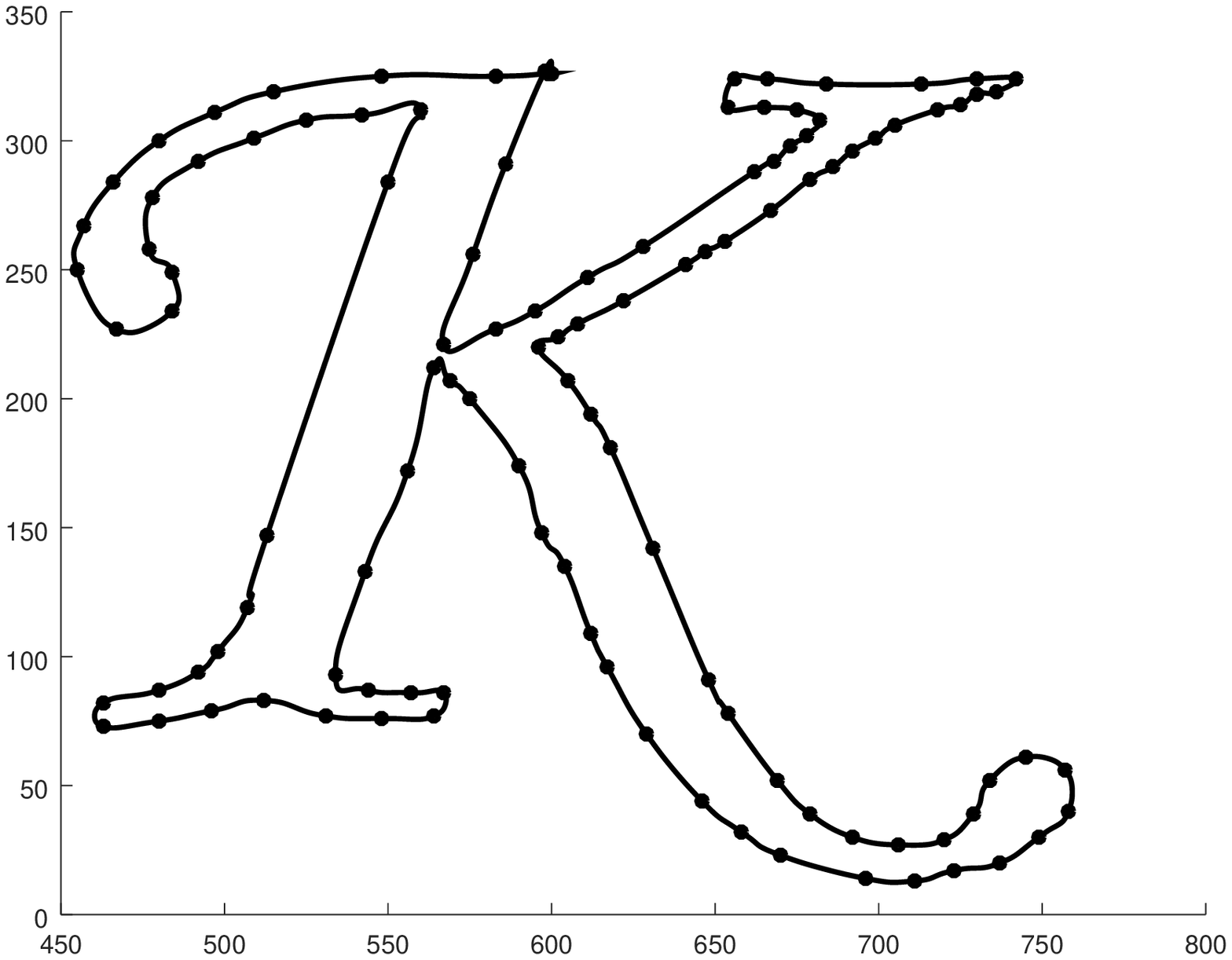}
        \caption{Approximated b-spline with 82\% points (error $\approx$ 11.16)
}
        \label{fig:gull}
    \end{subfigure}
    ~ 
    \begin{subfigure}[b]{0.3\textwidth}
        \includegraphics[width=\textwidth]{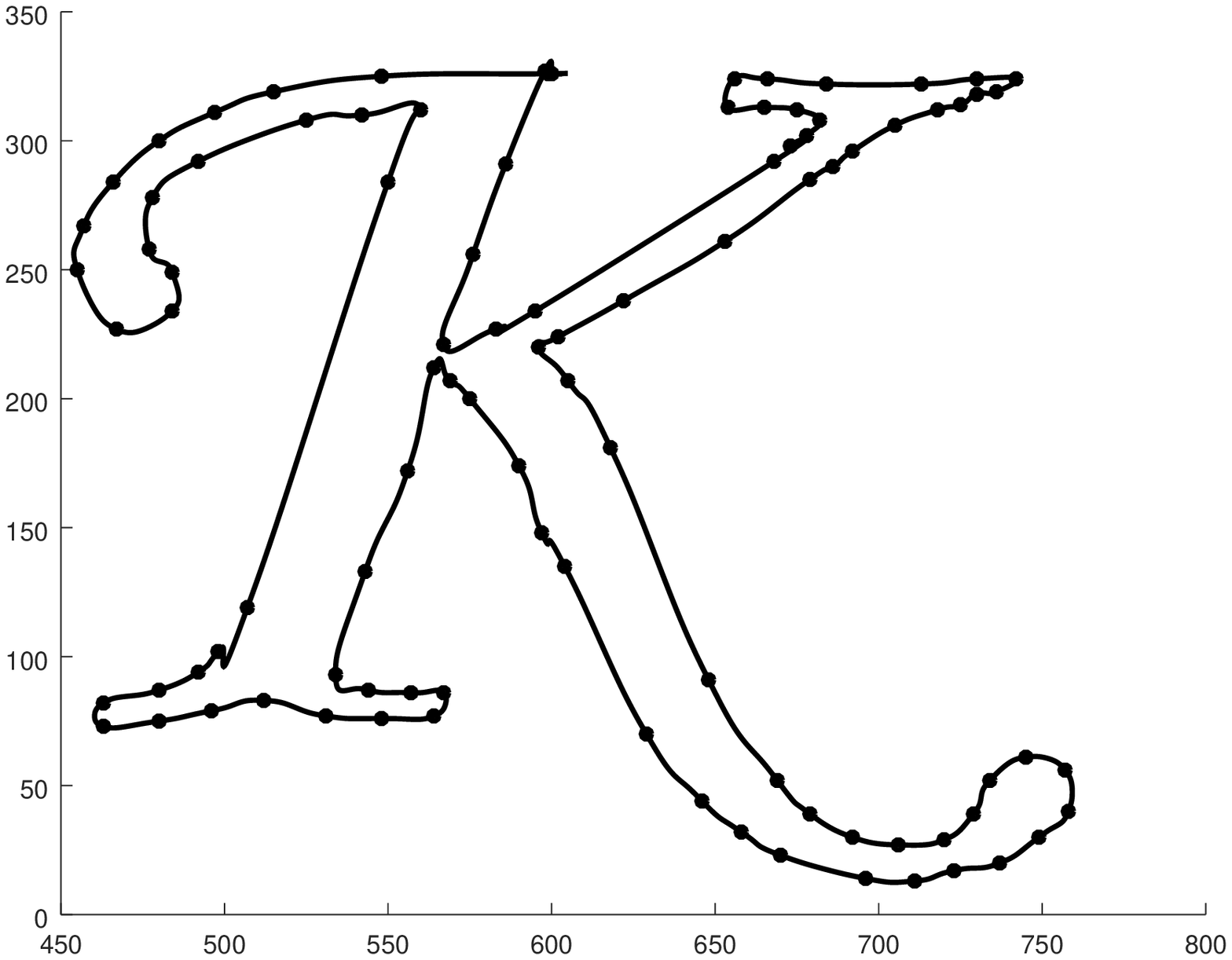}
        \caption{Approximated b-spline with 70\% points (error $\approx$ 20.89)
}
        \label{fig:tiger}
    \end{subfigure}
    ~ 
    \begin{subfigure}[b]{0.3\textwidth}
        \includegraphics[width=\textwidth]{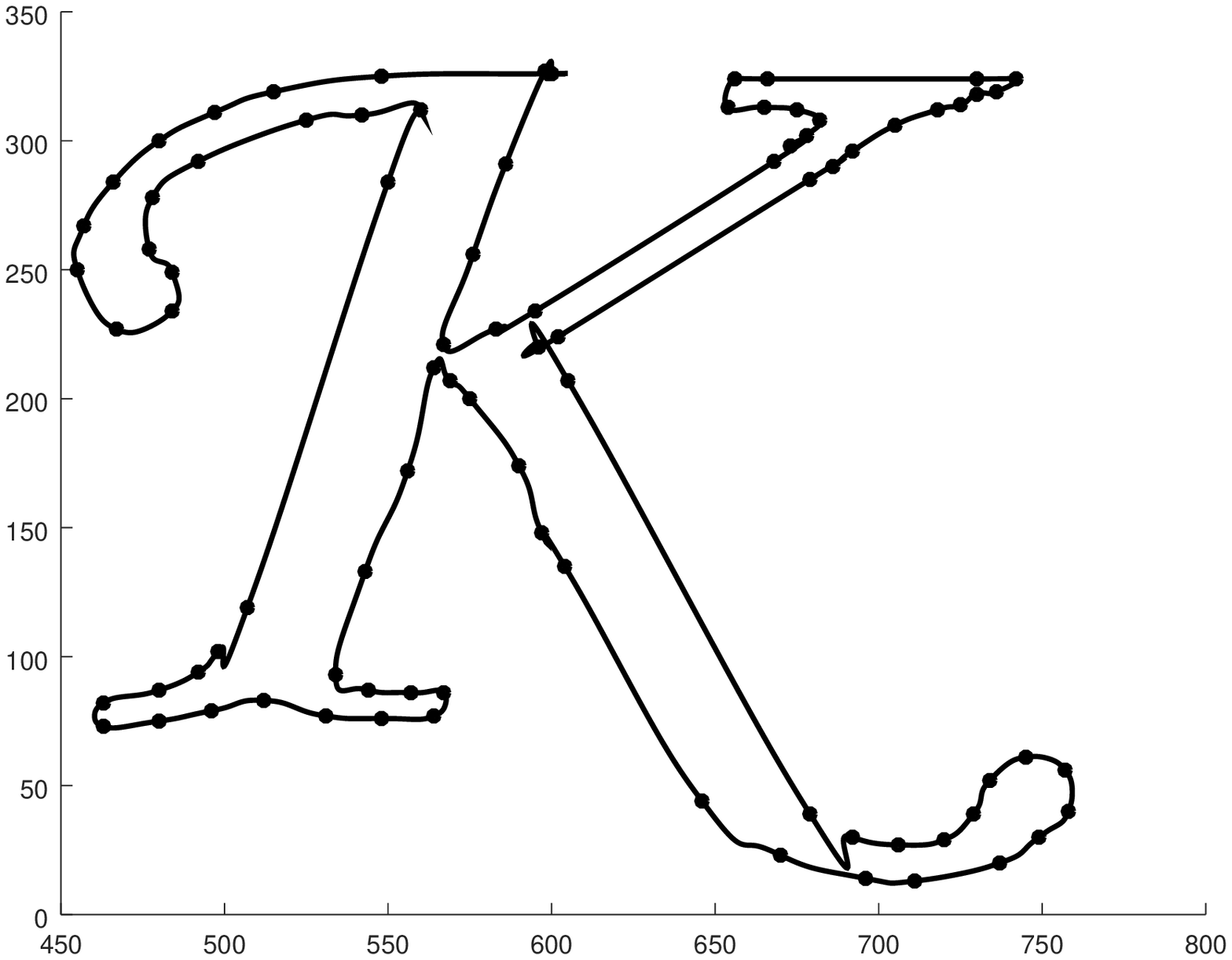}
        \caption{Approximated b-spline with 63\% points (error $\approx$ 366.34)}
        \label{fig:mouse}
    \end{subfigure}
    \caption{Experimentation 3}\label{res3_2dfig}
\end{figure}

\begin{figure}
    \centering
    \begin{subfigure}[b]{0.3\textwidth}
        \includegraphics[width=\textwidth]{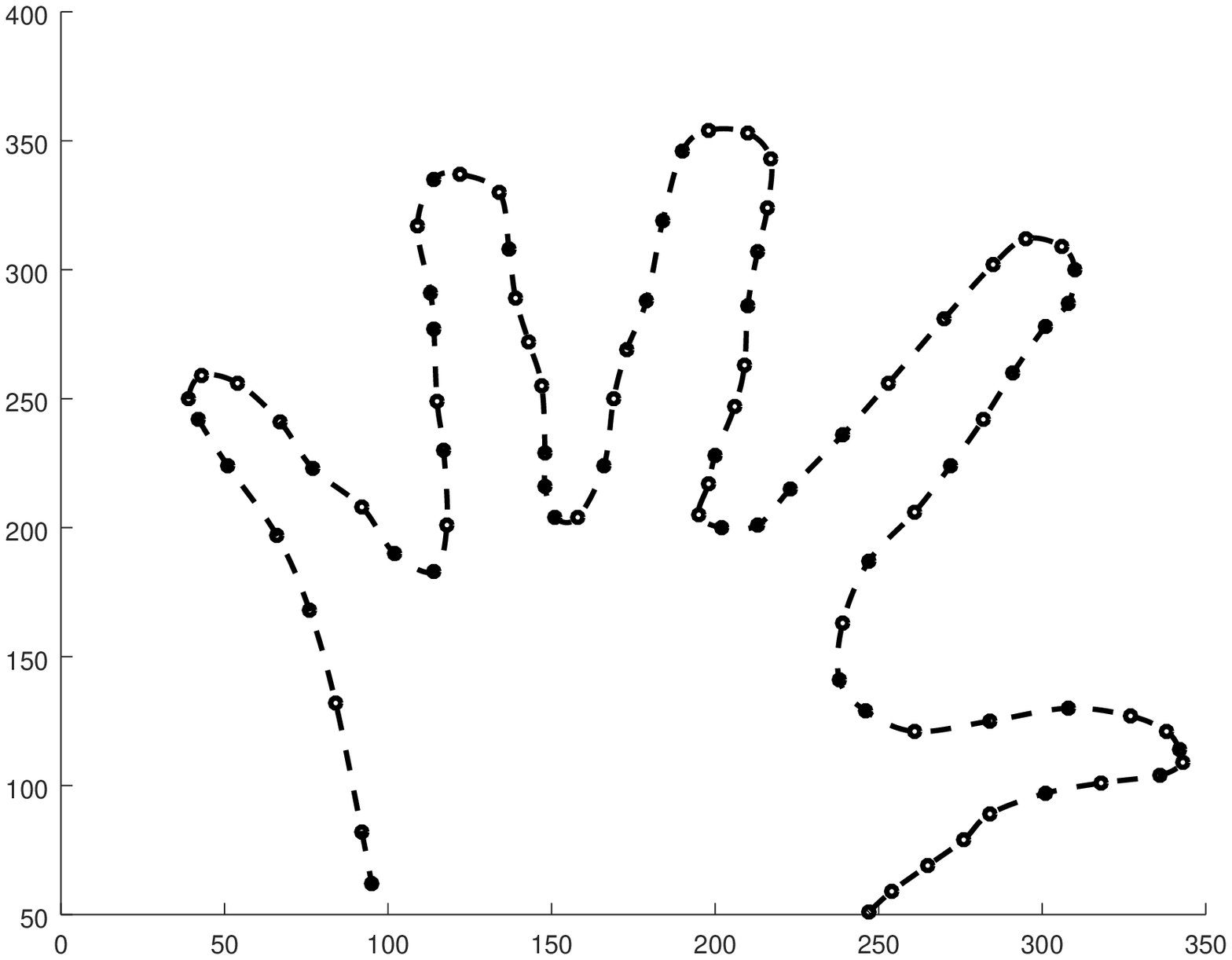}
        \caption{Interpolated Cardinal spline}
        \label{fig:gull}
    \end{subfigure}
    ~ 
    \begin{subfigure}[b]{0.3\textwidth}
        \includegraphics[width=\textwidth]{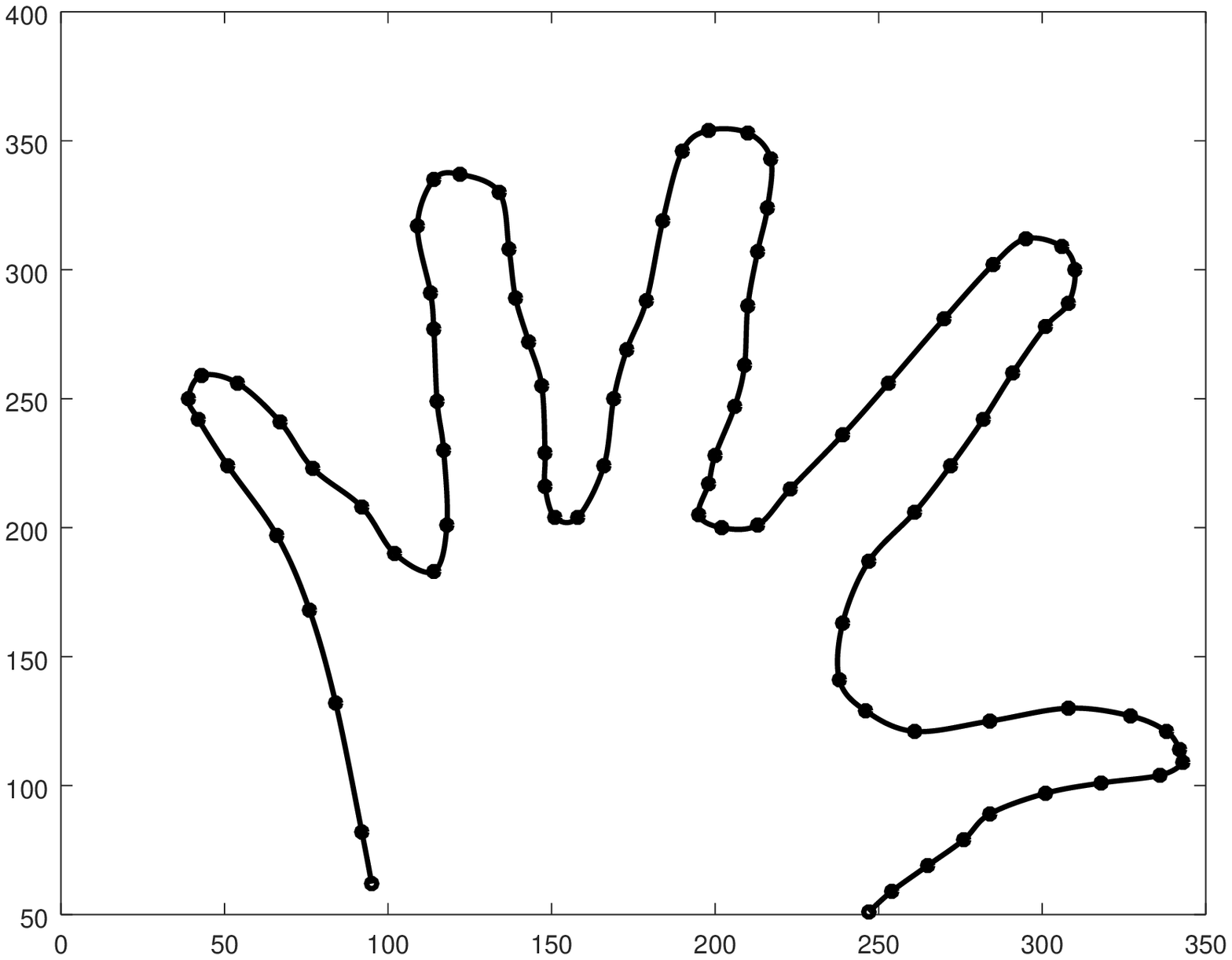}
        \caption{Piecewise cubic Bezier fit}
        \label{fig:tiger}
    \end{subfigure}
    ~ 
    \begin{subfigure}[b]{0.3\textwidth}
        \includegraphics[width=\textwidth]{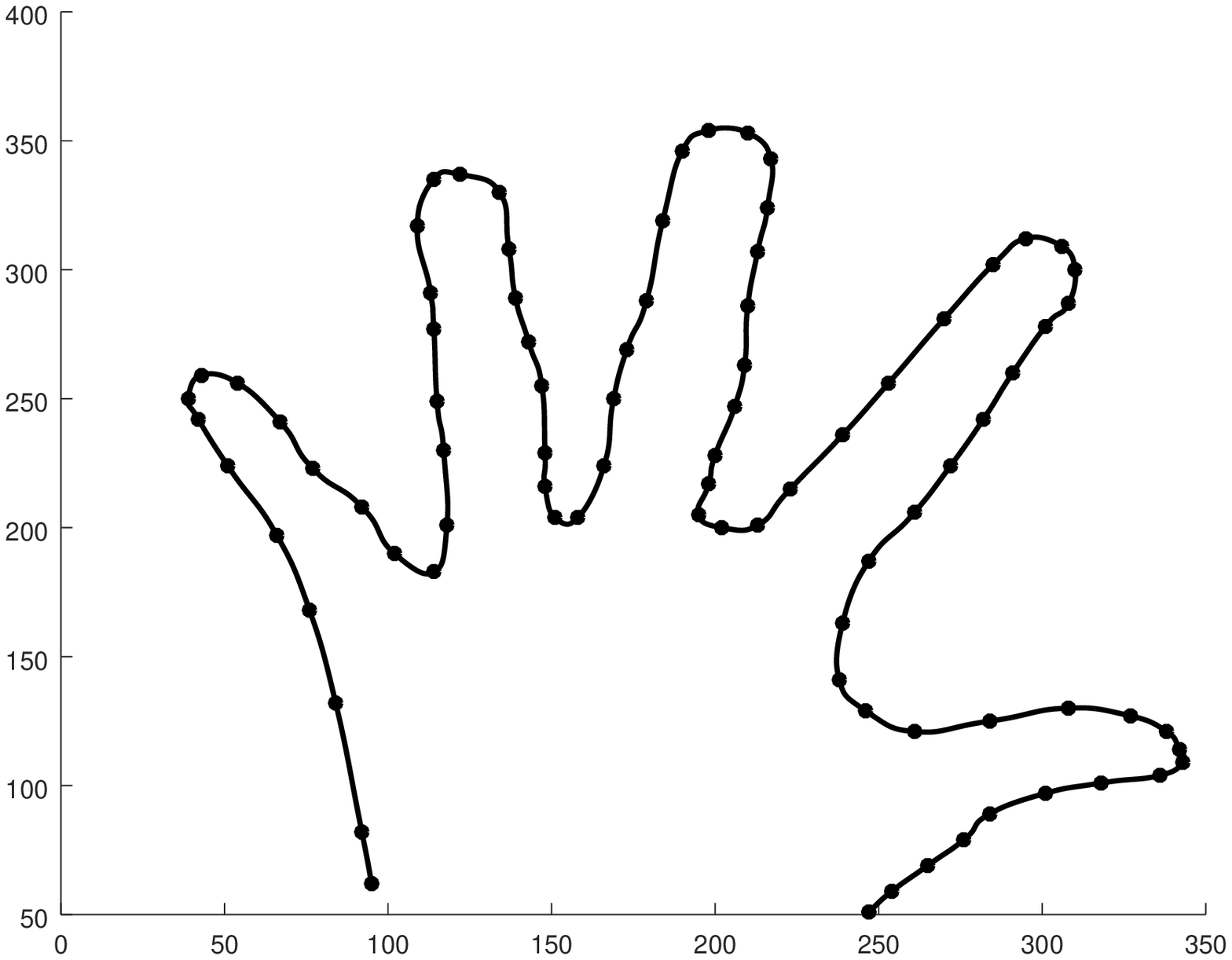}
        \caption{Cubic B-spline fit}
        \label{fig:mouse}
    \end{subfigure}
    \begin{subfigure}[b]{0.3\textwidth}
        \includegraphics[width=\textwidth]{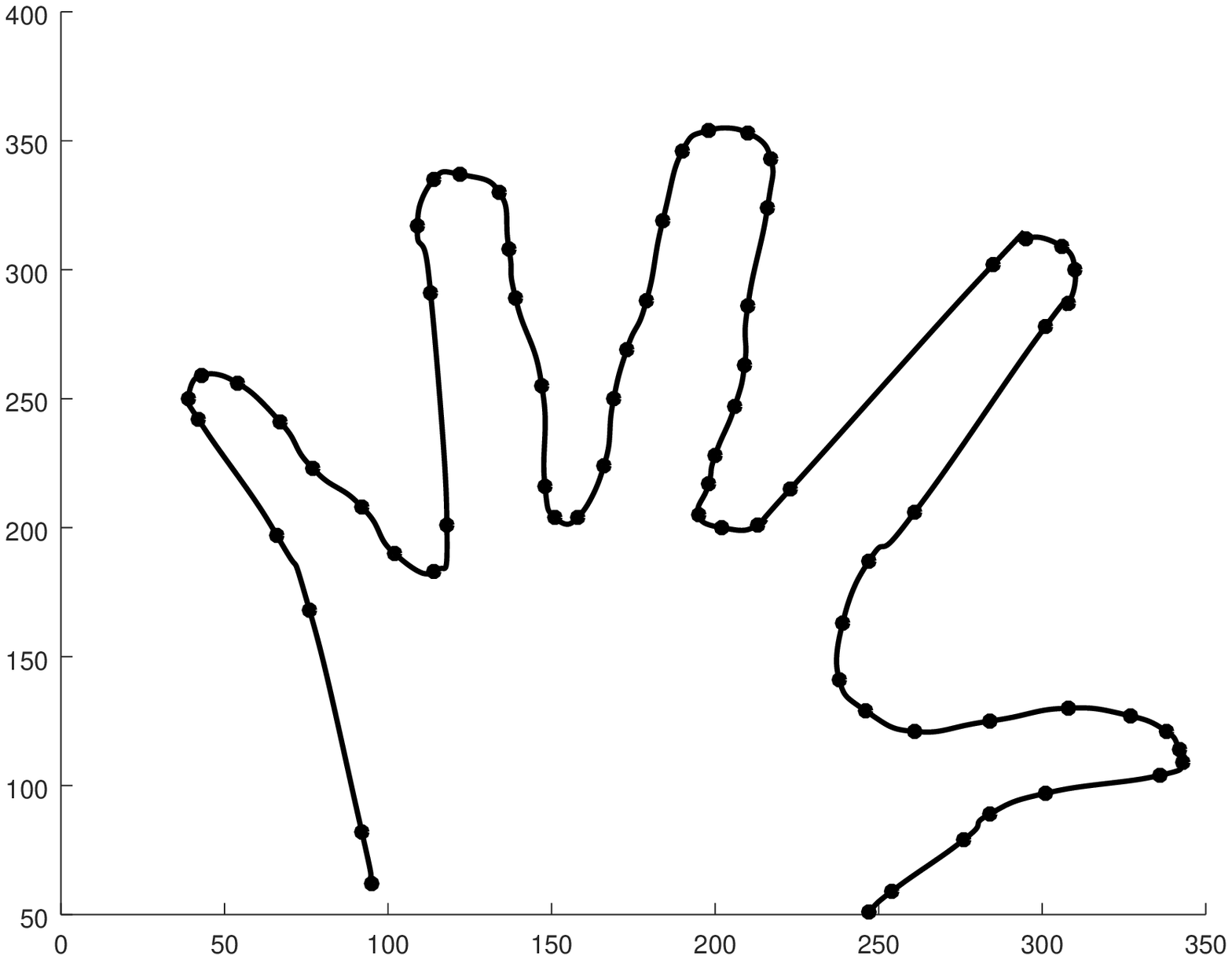}
        \caption{Approximated b-spline with 81\% points (error $\approx$ 5.66)
}
        \label{fig:gull}
    \end{subfigure}
    ~ 
    \begin{subfigure}[b]{0.3\textwidth}
        \includegraphics[width=\textwidth]{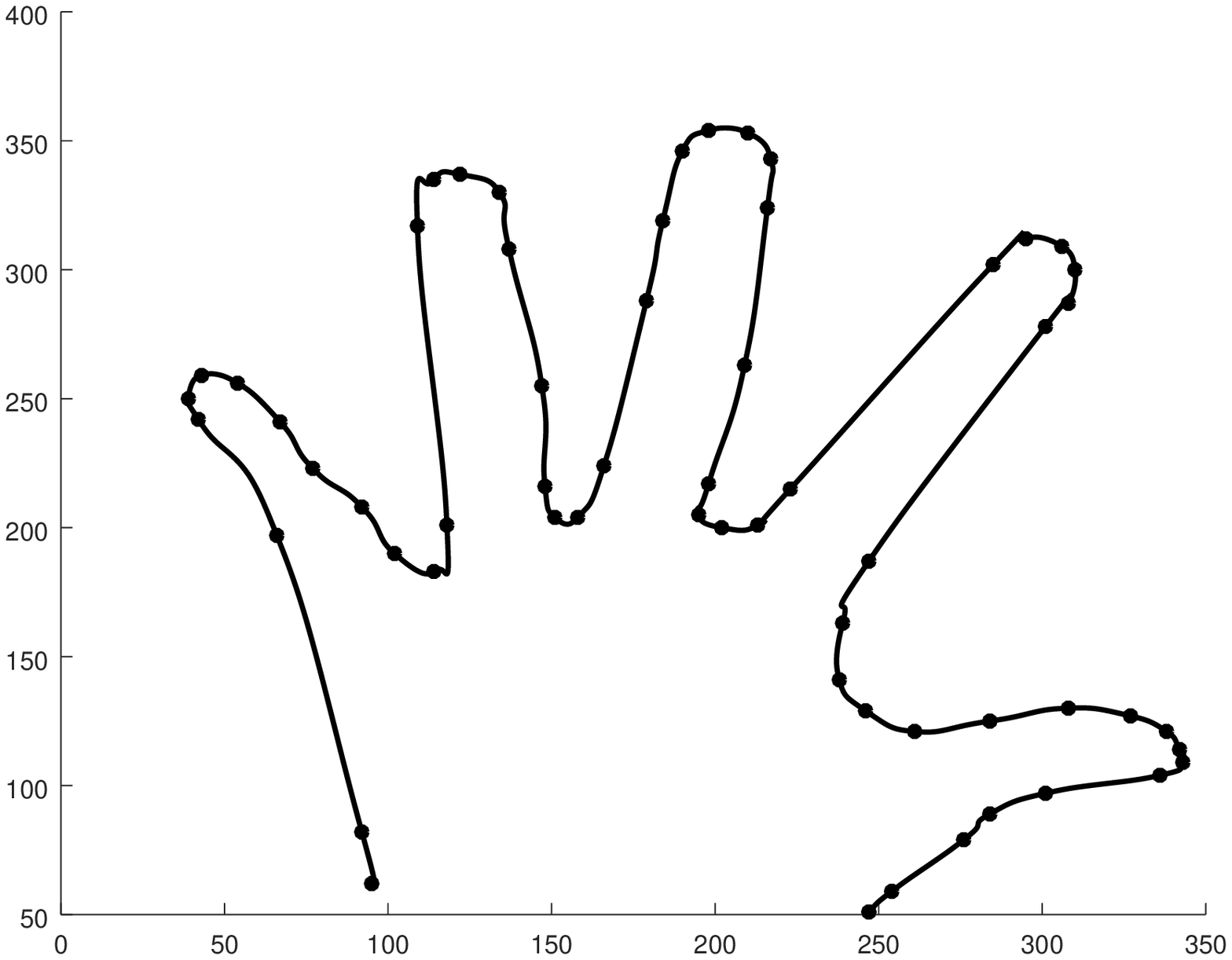}
        \caption{Approximated b-spline with 71\% points (error $\approx$ 67.24)
}
        \label{fig:tiger}
    \end{subfigure}
    ~ 
    \begin{subfigure}[b]{0.3\textwidth}
        \includegraphics[width=\textwidth]{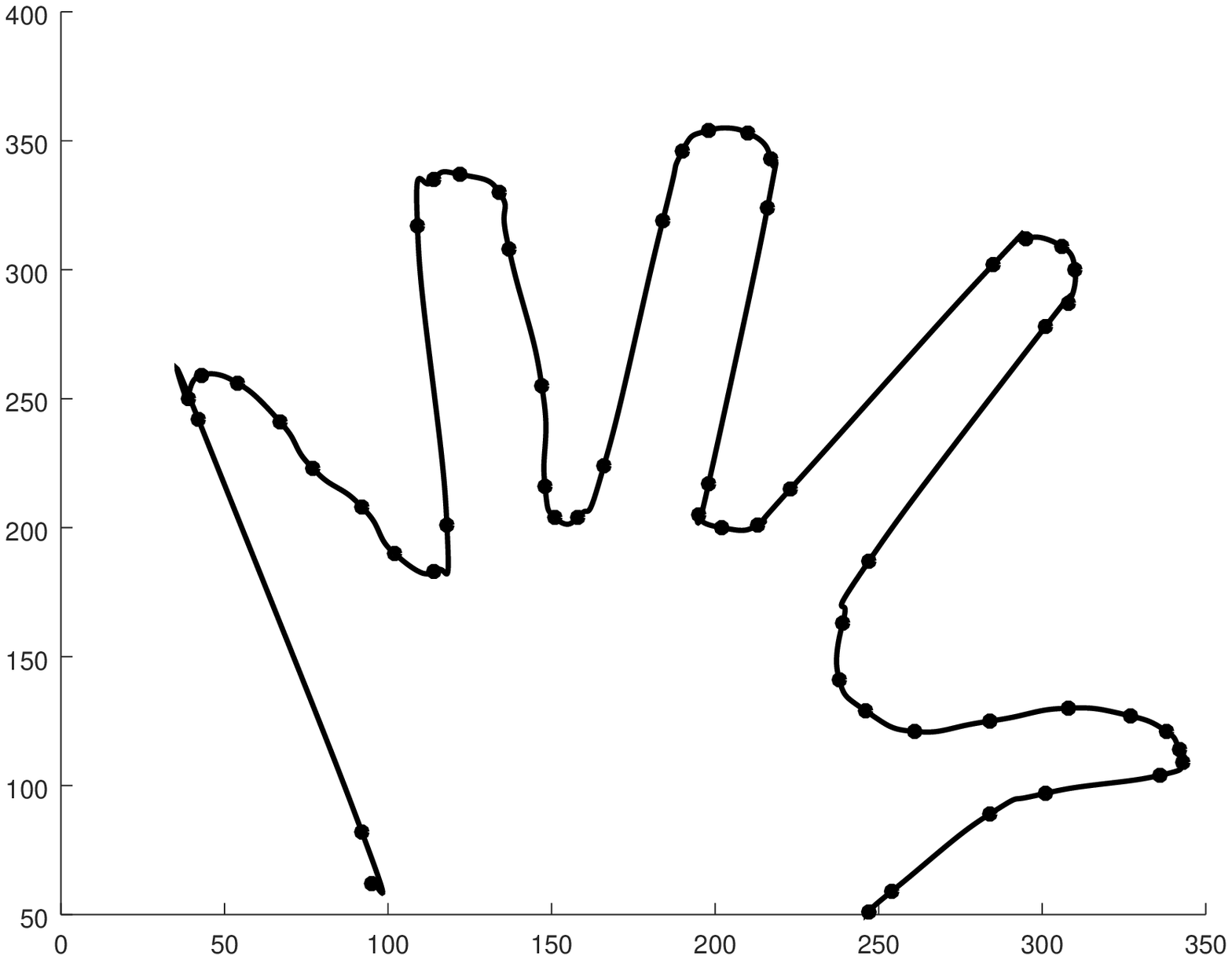}
        \caption{Approximated b-spline with 67\% points (error $\approx$ 323.64)}
        \label{fig:mouse}
    \end{subfigure}
    \caption{Experimentation 4}\label{res4_2dfig}
\end{figure}

Results of fitting b-spline curves to data points on coordinate plane are shown in Figures \ref{res1_2dfig} to \ref{res4_2dfig}. Each of the Figures contains Sub Figures a. to f. The Sub Figure a. in the Figures, show interpolation of the data points using a shape preserving, and visually pleasing, manner using cubic cardinal spline curve that preserves $C^{1}$ (or, tangent) continuity at junctions among consecutive segments, corresponding to the first step in FC2 algorithm. Piecewise cubic bezier curve, computed using control points for each piece, with end points and tangents, respective to the segments of cardinal spline, is shown in Sub Figure b, corresponding to the second step in FC2. Curvature continuous b-spline curve, using control points corresponding to the piecewise bezier curve in Sub Figure b., and based on the third step in FC2, is shown in Sub Figure c. Based on our algorithm presented in Section 2, approximating b-spline constructed using dominant points represented by fraction of the given points, in decreasing order of the fractions, and with corresponding least square errors, are shown in Sub Figures d., e., and f. The observations showed that, the least square error increase in exponents, with increased rate in lowering of the dominant points. Based on our experiments, 70\% of the points were observed to be dominant, for shapes given with dense data points, as in Figures \ref{res3_2dfig}, and \ref{res4_2dfig}.

\begin{figure}
    \centering
    \begin{subfigure}[b]{0.3\textwidth}
        \includegraphics[width=\textwidth]{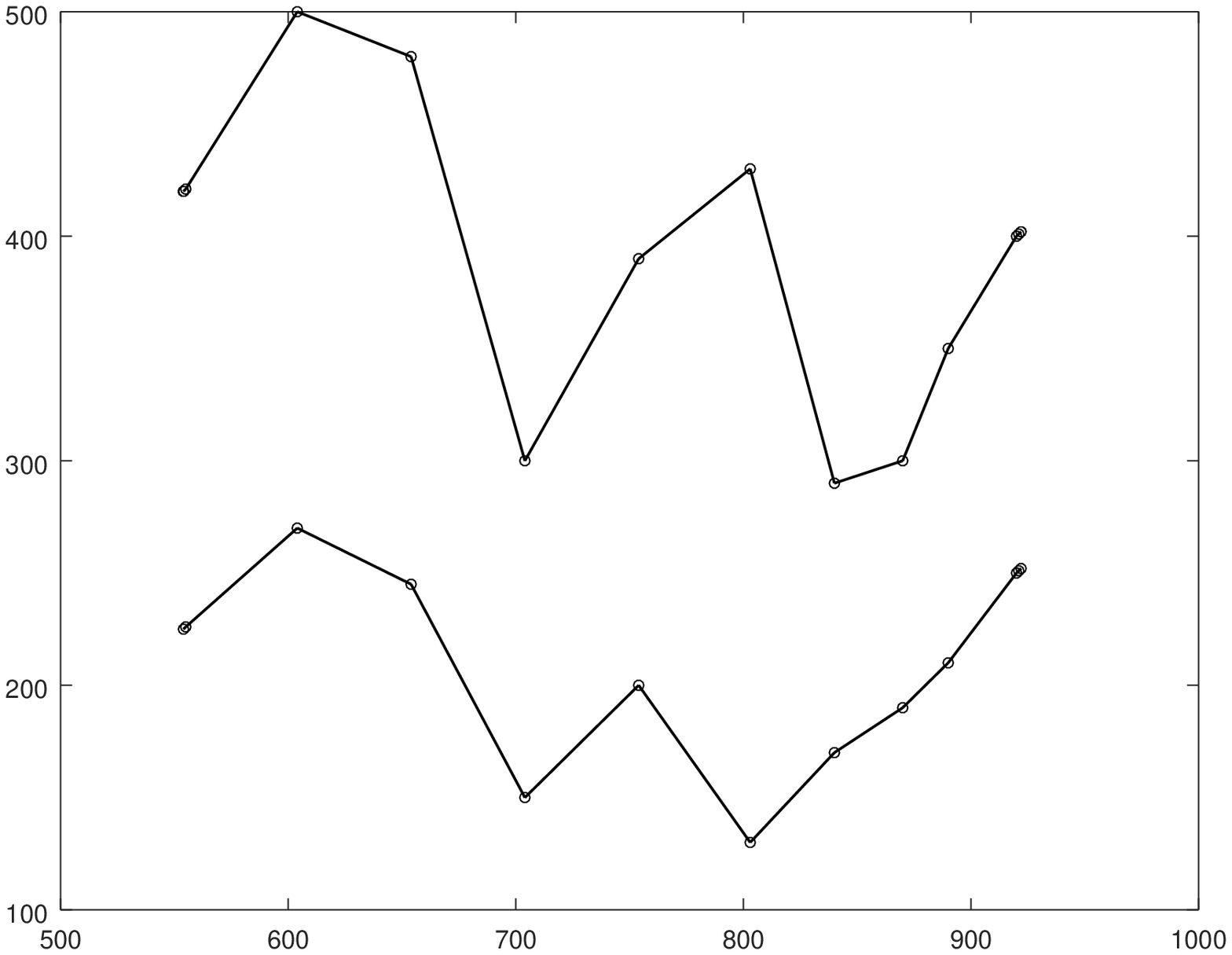}
        \caption{Input data points in YX and YZ coordinate planes.}
        \label{fig:gull}
    \end{subfigure}
    \begin{subfigure}[b]{0.3\textwidth}
        \includegraphics[width=\textwidth]{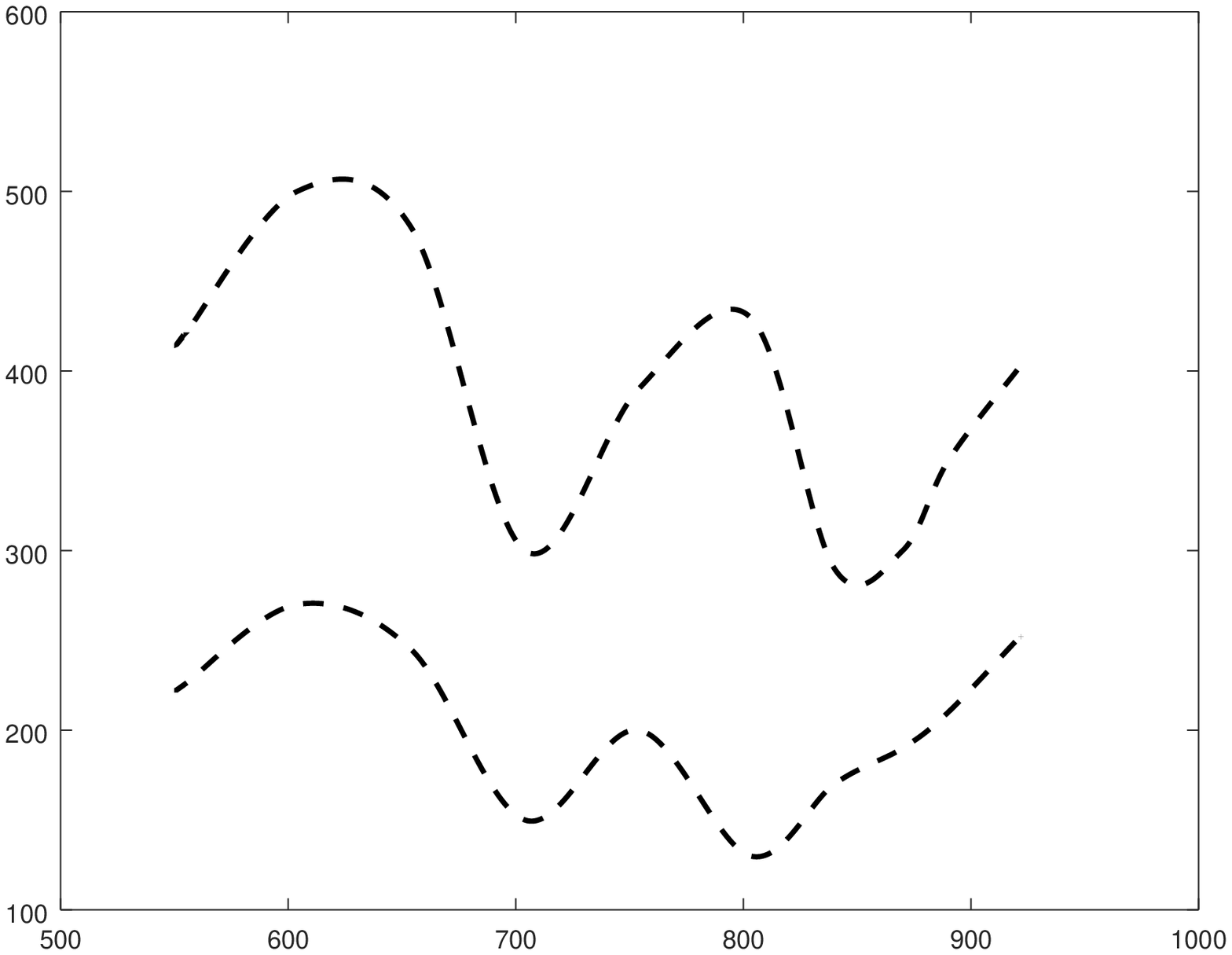}
        \caption{Interpolated Cardinal splines}
        \label{fig:gull}
    \end{subfigure}
    \begin{subfigure}[b]{0.3\textwidth}
        \includegraphics[width=\textwidth]{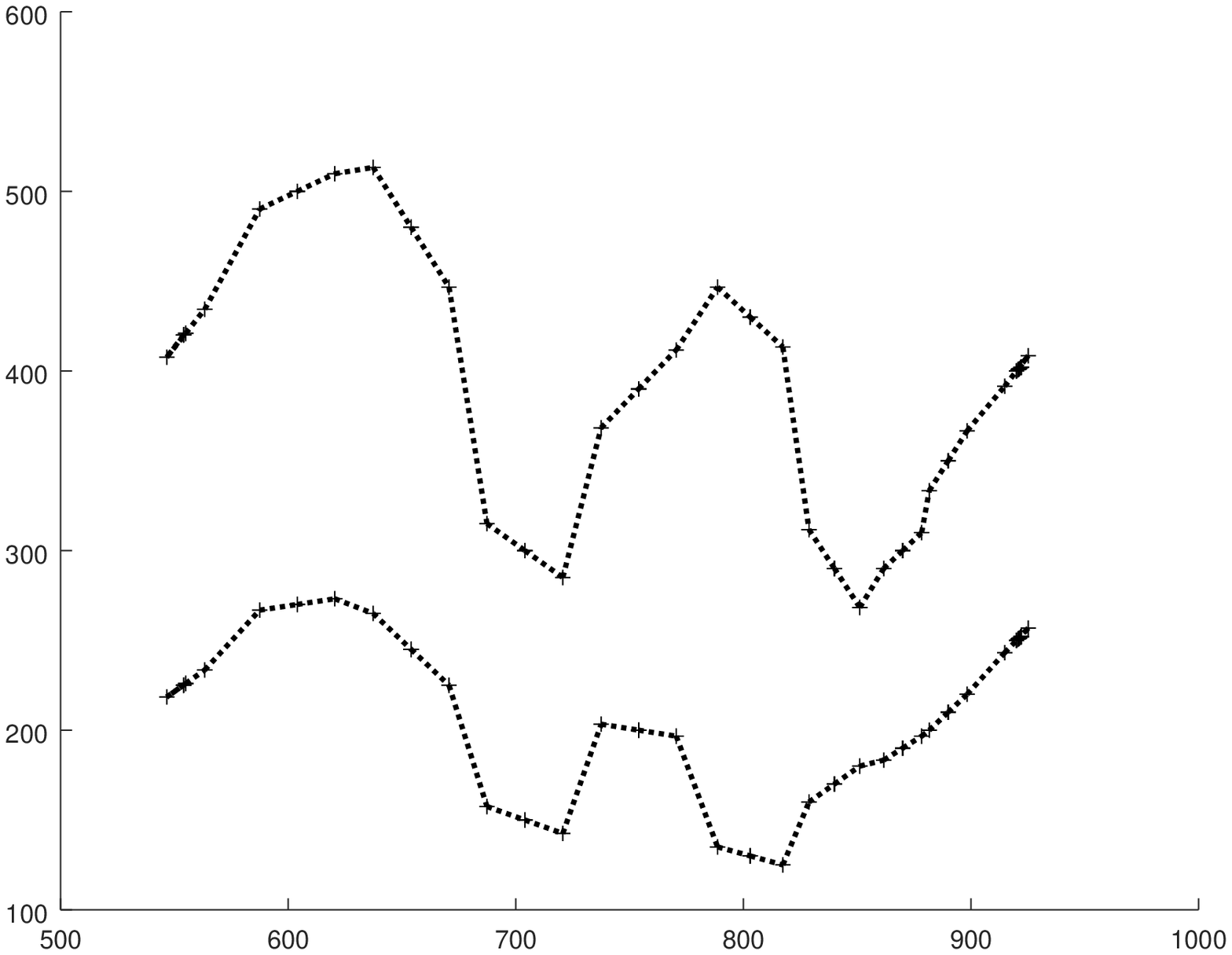}
        \caption{Control polygons for piecewise Bezier curves}
        \label{fig:gull}
    \end{subfigure}
    ~ 
    \begin{subfigure}[b]{0.3\textwidth}
        \includegraphics[width=\textwidth]{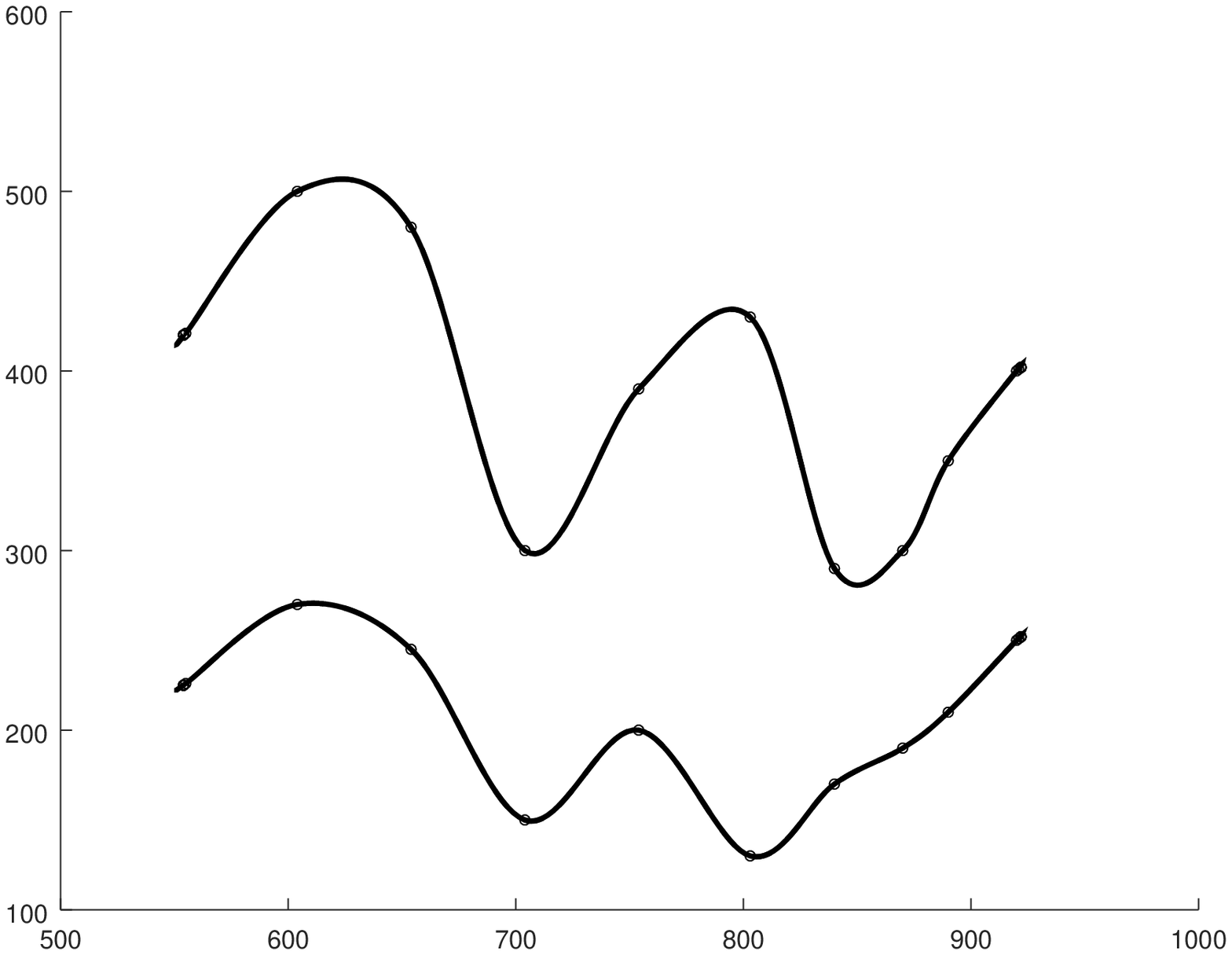}
        \caption{Piecewise cubic Bezier fits}
        \label{fig:tiger}
    \end{subfigure}
    ~ 
    \begin{subfigure}[b]{0.3\textwidth}
        \includegraphics[width=\textwidth]{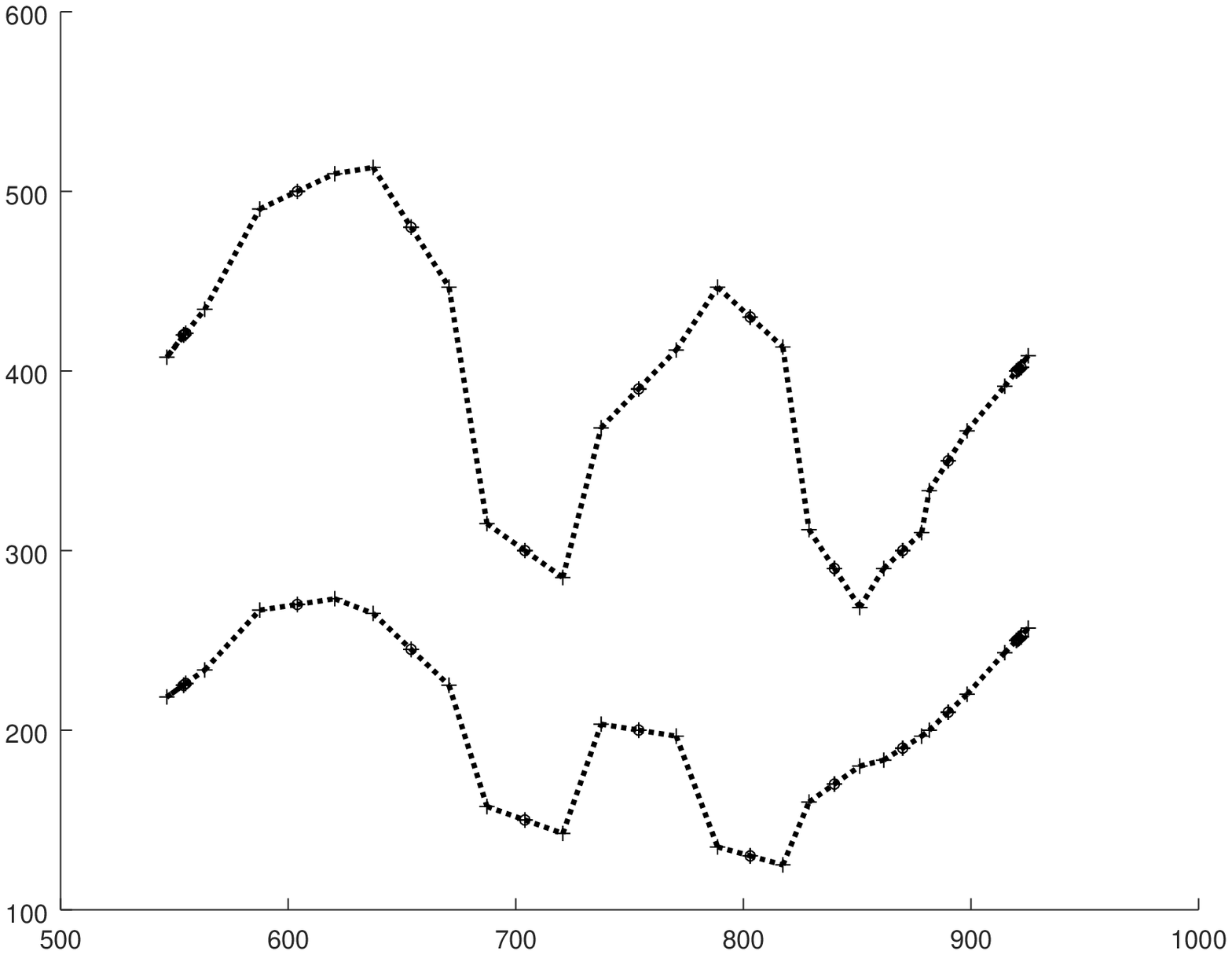}
        \caption{Merged control polygons on coordinate planes, for computation of the space curve.}
        \label{fig:mouse}
    \end{subfigure}
    \begin{subfigure}[b]{0.3\textwidth}
        \includegraphics[width=\textwidth]{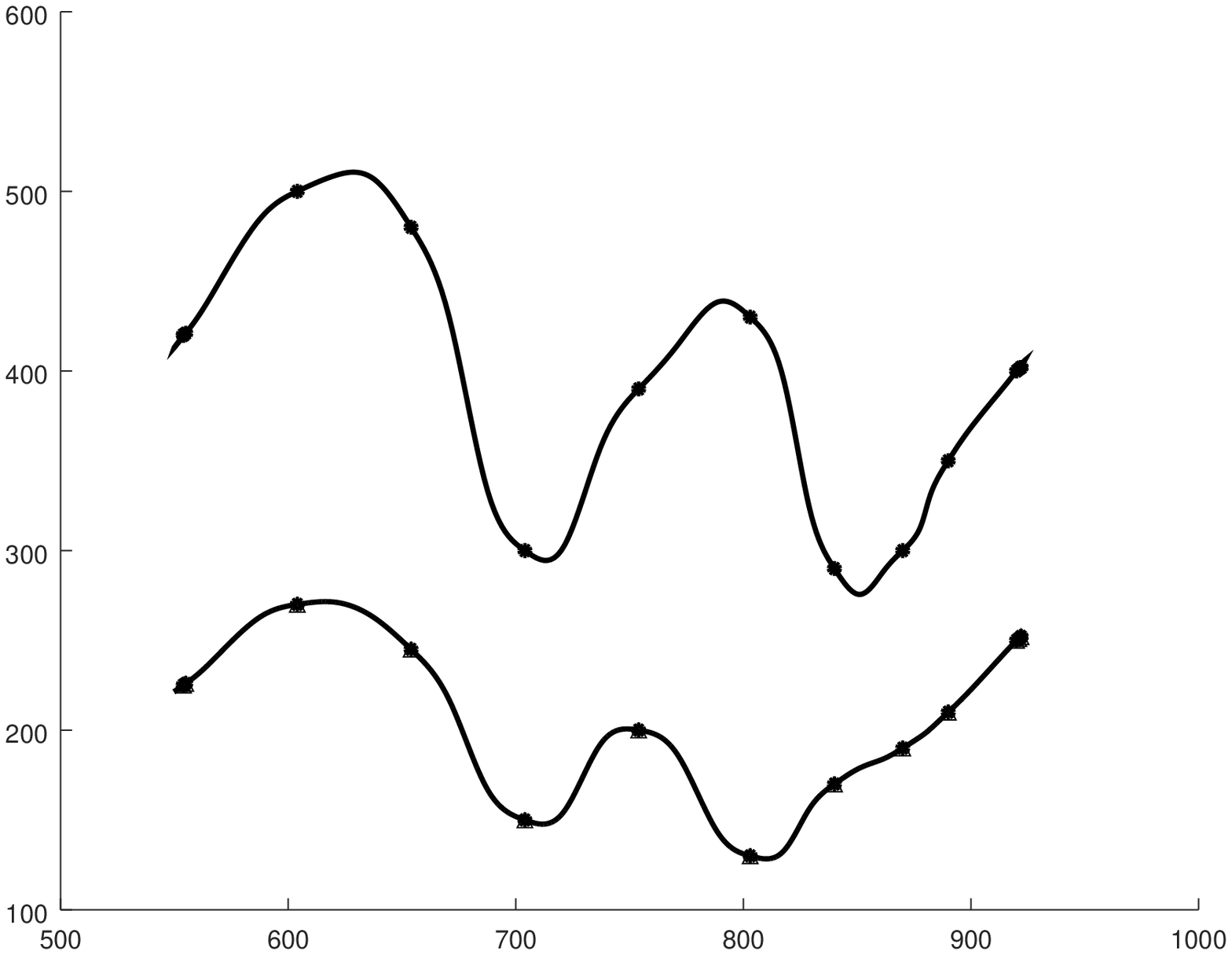}
        \caption{Cubic B-splines corresponding to merged control polygons
}
        \label{fig:gull}
    \end{subfigure}

    \caption{Fitting B-spline curve to data points in three dimensional space. }\label{res5_3dfig}
\end{figure}

We have experimented the present approach, on fitting b-spline curve, to data point coordinates in the three dimensional space, on multiple data sets prepared by us.   Results of experiment, on an example data, is shown in Figure \ref{res5_3dfig}. Sub Figure \ref{res5_3dfig}.a, shows the data points in the $\mathbb{R}^{3}$ space, reduced to the data points in the corresponding coordinate planes YX and YZ.  Curve fitting using FC2 are applied to the data points separately in each of the two coordinate planes. Interpolation using cardinal splines, computation of control points in corresponding planes, and piecewise cubic beziers fitted to the data points, are shown in Figure \ref{res5_3dfig}.b, \ref{res5_3dfig}.c, \ref{res5_3dfig}.d, respectively. Merging of control points in the two coordinate planes, are considered corresponding to control points of the respective pieces of the bezier curves selected from Sub Figure \ref{res5_3dfig}.c, using the first two steps of FC3 algorithm. The control points on the coordinate planes after merging of coordinates, based on the next step, are shown in Sub Figure \ref{res5_3dfig}.e. Based on this step in FC3, the control points with the merged coordinates are used to produce the points in the higher dimensional space. Finally, the control points thus obtained are then used to compute the cubic b-spline curve approximating the data points in the three dimensional space, in the last step of FC3. The projection of the space curve, on the corresponding coordinate planes, is shown in Sub Figure \ref{res5_3dfig}.f.

\section{Conclusions}
We have presented an approach for fitting B-spline curves to data points in three dimensional space. 
The present solution can approximate data points with cubic b-spline curves within a given bound of error in the least square sense.  The data point that is next to a present is required to be explicit at input. This fact of the input is used for a shape preserving and visually pleasing interpolation, free from wiggles, as a prerequisite in the present scheme. Fitting to unordered data, based on assumptions on parameters implicitly, with any knowledge on its natural evolution, is not possible in our approach at present, and is a shortcoming.  The solution presented though basic, can be both refined for increased continuity or quality of fit, and extended to higher dimensions, based on modifications to knot vectors and merging of control points from dependent lower dimension spaces, respectively.
Our work on dominant points is inline with present work on b-splines to fit higher dimensional data.
The approach presented for curves, automatically extends to representations for surfaces that may be of bi-parametric and ruled.

\bibliographystyle{elsarticle-num}


\end{document}